\documentclass[structabstract]{aa}  
\usepackage{graphicx}
\usepackage{txfonts}
\begin{document}
\title{The close circumstellar environment of Betelgeuse}

\subtitle{II. Diffraction-limited spectro-imaging from 7.76 to 19.50\,$\mu$m with VLT/VISIR\thanks{Based on observations made with ESO telescopes at Paranal Observatory, under ESO DDT program 286.D-5007(A).}}
\titlerunning{The close circumstellar environment of Betelgeuse from 7.76 to 19.50\,$\mu$m}

\author{
P. Kervella \inst{1}
\and
G. Perrin \inst{1}
\and
A. Chiavassa \inst{2}
\and
S. T. Ridgway \inst{3}
\and
J. Cami\inst{4,5}
\and
X. Haubois\inst{6}
\and
T. Verhoelst \inst{7}
 }
\offprints{Pierre Kervella}
\mail{pierre.kervella@obspm.fr}

\institute{
LESIA, Observatoire de Paris, CNRS\,UMR\,8109, UPMC, Universit\'e Paris Diderot, 5 place Jules Janssen, 92195 Meudon, France
\and
Institut d'Astronomie et d'Astrophysique, Universit\'e Libre de Bruxelles, CP\,226, Boulevard du Triomphe, B-1050 Brussels, Belgium
\and
National Optical Astronomy Observatories, 950 North Cherry Avenue, Tucson, AZ 85719, USA
\and
Physics and Astronomy Dept, University of Western Ontario, London ON N6A 3K7, Canada
\and
SETI Institute, 189 Bernardo Ave, Suite 100, Mountain View, CA 94043, USA
\and
Instituto de Astronomia, Geof'sica e Ci\^encias Atmosf\'ericas, Universidade de S\~ao Paulo, Rua do Mat\~ao 1226, Cidade Universit\'aria, 05508-900 S\~ao Paulo, SP, Brazil
\and
Instituut voor Sterrenkunde, K. U. Leuven, Celestijnenlaan 200D, B-3001 Leuven, Belgium
}

\date{Received 25 March 2011; Accepted 30 May 2011}

  \abstract
{Mass-loss occurring in red supergiants (RSGs) is a major contributor to the enrichment of the interstellar medium in dust and molecules. The physical mechanism of this mass loss is however relatively poorly known. Betelgeuse is the nearest RSG, and as such a prime object for high angular resolution observations of its surface (by interferometry) and close circumstellar environment.}
{The goal of our program is to understand how the material expelled from Betelgeuse is transported from its surface to the interstellar medium, and how it evolves chemically in this process.}
{We obtained diffraction-limited images of Betelgeuse and a calibrator (Aldebaran) in six filters in the $N$ band (7.76 to 12.81\,$\mu$m) and two filters in the $Q$ band (17.65 and 19.50\,$\mu$m), using the VLT/VISIR instrument.}
{Our images show a bright, extended and complex circumstellar envelope at all wavelengths. It is particularly prominent longwards of $\approx$9-10\,$\mu$m, pointing at the presence of O-rich dust, such as silicates or alumina. A partial circular shell is observed between 0.5 and 1.0$\arcsec$ from the star, and could correspond to the inner radius of the dust envelope. Several knots and filamentary structures are identified in the nebula. One of the knots, located at a distance of $0.9\arcsec$ west of the star, is particularly bright and compact.}
{The circumstellar envelope around Betelgeuse extends at least up to several tens of stellar radii. Its relatively high degree of clumpiness indicates an inhomogeneous spatial distribution of the material lost by the star. Its extension corresponds to an important intermediate scale, where most of the dust is probably formed, between the hot and compact gaseous envelope observed previously in the near infrared and the interstellar medium.}


\maketitle
%

\section{Introduction}

Several mysteries remain to be solved for evolved supergiant stars, among which the structure of their convection, the mechanism of their mass loss (Levesque~\cite{levesque10}) and the way dust forms in their environment (Verhoelst et al.~\cite{verhoelst09}). Josselin et~al.~(\cite{josselin00}) showed that the dust mass-loss rate is not correlated with luminosity, and that the molecular gas-to-dust ratio shows a very large scatter, which is higher than what is observed in asymptotic giant branch (AGB) stars. The mass loss mechanism of AGB stars, based on pulsations and radiation pressure on dust grains, is unlikely to be applicable to red supergiant (RSG) stars because they are irregular variables with small amplitudes. The mass loss mechanism in RSGs could be related to the turbulent pressure generated by convective motions, which, combined with radiative pressure on molecular lines, may levitate the gas (Josselin $\&$ Plez~\cite{josselin07}). The red supergiant Betelgeuse (spectral type M1-M2 Ia-Ib, $T_{\rm eff}=3650$\,K, Levesque et al.~\cite{levesque05}) is known to have a peculiar convective pattern on its surface characterized by small to large convective cells (up to more than one stellar radius in size) with supersonic velocities and shocks (Chiavassa et~al.~\cite{chiavassa10}). Moreover, the magnetic field discovered on Betelgeuse (Auri{\`e}re et~al.~\cite{auriere10}) may also contribute to the mass loss through Alfv\'en waves (see e.g. Hartmann et~al.~\cite{Hartmann84}, Pijpers et~al.~\cite{Pijpers89}, Cuntz et~al.~\cite{cuntz97}, Airapetian et~al.~\cite{airapetian10}). Progress on our understanding of this intricate combination of physical phenomena requires high angular resolution studies of the surface of the star using long-baseline interferometry (e.g. Haubois et~al.~\cite{haubois09}), and its circumstellar environment (hereafter CSE).

Evidence for mass-loss is well established in late-type stars with dust detected tens and hundreds of stellar radii away from the stellar surface. However, the way dust is created still remains a partial mystery, although progress has been made recently on both the observational and theoretical sides. The concept of MOLsphere was first hypothesized spectroscopically in the environment of the red supergiants Betelgeuse and $\mu$\,Cep (Tsuji~\cite{tsuji00}). Spectro-Interferometric observations confirmed this hypothesis and established some characteristics of the MOLsphere such as its height and composition (Perrin et~al.~\cite{perrin04}, Verhoelst et~al.~\cite{verhoelst06}, Perrin et~al.~Ê\cite{perrin07}, see also Tsuji~\cite{tsuji06} and Schuller et al.~\cite{schuller04}).
This scenario is a candidate to explain how dust forms in red supergiants, but does not provide rock-solid arguments to explain how the MOLsphere can be formed and levitated above the supergiant photosphere. One hypothesis is that convection cells detected by Chiavassa et~al.~(\cite{chiavassa10}) are at the root of mass transfer from the star surface to its upper atmosphere (Ohnaka et~al.~\cite{ohnaka09}), and that the expelled material condenses as dust within the MOLsphere and above.

In January 2009, Kervella et~al.~(\cite{kervella09}, hereafter ``Paper~I") obtained adaptive optics (AO) images of \object{Betelgeuse} with NACO, taking advantage of the cube mode of the CONICA camera to record separately a large number of short-exposure frames. With a selective imaging approach, they obtained diffraction-limited images over the spectral range $1.04-2.17\,\mu$m in 10 narrow-band filters. These AO images show that the inner CSE of Betelgeuse has a complex and irregular structure, with in particular a bright ``plume" extending in the southwestern quadrant up to a radius of at leasy six times the photosphere. These observations allowed the probable identification of the CN molecule, which is relatively common in the envelopes of evolved post-RGB stars (e.g. Cherchneff~\cite{cherchneff06}). We have here a first piece of evidence for the mass-loss puzzle of Betelgeuse. A next step requires detection of the dust itself, as an indicator of the location of formation. Considering the temperature at which dust is supposed to form, this requires thermal infrared imaging. This is the object of the new VISIR observations in the $N$ and $Q$ bands reported in the present article. The observations and raw data processing are presented in Sect.~\ref{observations}, their analysis in Sect.~\ref{analysis}, and we discuss our results in Sect.~\ref{discussion}.

\section{Observations \label{observations}}

\subsection{Instrumental setup and data acquisition}

For our observations of \object{Betelgeuse} ($\alpha$\,Ori, \object{HD 39801}, \object{HR 2061}) and its point spread function (hereafter PSF) calibrator \object{Aldebaran} ($\alpha$\,Tau, \object{HD 29139}, \object{HR 1457}), we used the VISIR instrument (Lagage et~al.~\cite{lagage04}), installed at the Cassegrain focus of the Melipal telescope (UT3) of ESO's Very Large Telescope (Paranal, Chile). VISIR is a mid-infrared (hereafter MIR) imager and slit spectrometer, that can in principle reach a very high angular resolution, thanks to the 8\,m diameter of the telescope. However, under standard conditions at Paranal (median seeing of 0.8$\arcsec$ at 0.5\,$\mu$m), the 8\,m telescope is not diffraction limited in the MIR (seeing $\approx 0.4\arcsec$ vs. 0.3$\arcsec$ diffraction). A few moving speckles and tip-tilt usually degrade the quality of the image (see e.g. Tokovinin, Sarazin \& Smette~\cite{tokovinin07}). To overcome this limitation, a specific mode of the instrument, the BURST mode (Doucet et~al.~\cite{doucet07}), has been implemented to give the possibility to record separately a large number of very short exposures ($\Delta t \lesssim 50$\,ms), in order to freeze the turbulence (see e.g. Law et al.~\cite{law06}, Law et al.~\cite{law09}, and references therein). To obtain diffraction limited images of the CSE of Betelgeuse, we therefore selected this instrument mode.

\begin{table}
\caption{Log of the VISIR observations of Betelgeuse and Aldebaran. } 
\label{visir_log}
\begin{tabular}{lccrrcc}
\hline \hline
Star & MJD$^*$ & Filter & NDIT & DIT & $\theta$\ & AM \\
 &  & & & (ms) & $(\arcsec)$ & \\
\hline
\noalign{\smallskip}
$\alpha$\,Tau & 12.2877 & J7.9 & 1600 & 40 & 0.70 & 1.38 \\
$\alpha$\,Tau & 12.2897 & PAH1 & 6400 & 10 & 0.69 & 1.39 \\
$\alpha$\,Tau & 12.2917 & SIV & 4000 & 16 & 0.75 & 1.40 \\
$\alpha$\,Tau & 12.2936 & PAH2 & 8000 & 8 & 0.63 & 1.40 \\
$\alpha$\,Tau & 12.2957 & SiC & 5120 & 12.5 & 0.78 & 1.41 \\
$\alpha$\,Tau & 12.2977 & NeII & 5120 & 12.5 & 1.02 & 1.42 \\
$\alpha$\,Tau & 12.3005 & Q1 & 10240 & 12.5 & 1.08 & 1.43 \\
$\alpha$\,Tau & 12.3039 & Q3 & 16000 & 8 & 1.27 & 1.44 \\
\hline             
\noalign{\smallskip}             
$\alpha$\,Ori & 12.3136 & J7.9 & 1800 & 40 & 1.89 & 1.18 \\
$\alpha$\,Ori & 12.3166 & PAH1 & 14400 & 10 & 1.51 & 1.19 \\
$\alpha$\,Ori & 12.3203 & SIV & 9000 & 16 & 1.59 & 1.19 \\
$\alpha$\,Ori & 12.3239 & PAH2 & 16000 & 8 & 1.40 & 1.19 \\
$\alpha$\,Ori & 12.3274 & SiC & 11520 & 12.5 & 1.66 & 1.20 \\
$\alpha$\,Ori & 12.3309 & NEII & 11520 & 12.5 & 1.91 & 1.20 \\
$\alpha$\,Ori & 12.3337 & Q1 & 6080 & 12.5 & 1.75 & 1.21 \\
$\alpha$\,Ori & 12.3357 & Q3 & 8000 & 8 & 1.70 & 1.21 \\
\hline
\end{tabular}
\tablefoot{MJD$^*$ is the modified Julian date of the start the exposures on the target minus 55\,500. The Detector Integration Time (DIT) is the exposure time for one BURST image, $\theta$ the seeing in the visible ($\lambda=0.5\,\mu$m) measured by the observatory DIMM sensor, and AM the airmass of the observation.}
\end{table}

\begin{figure}[]
\includegraphics[width=\hsize]{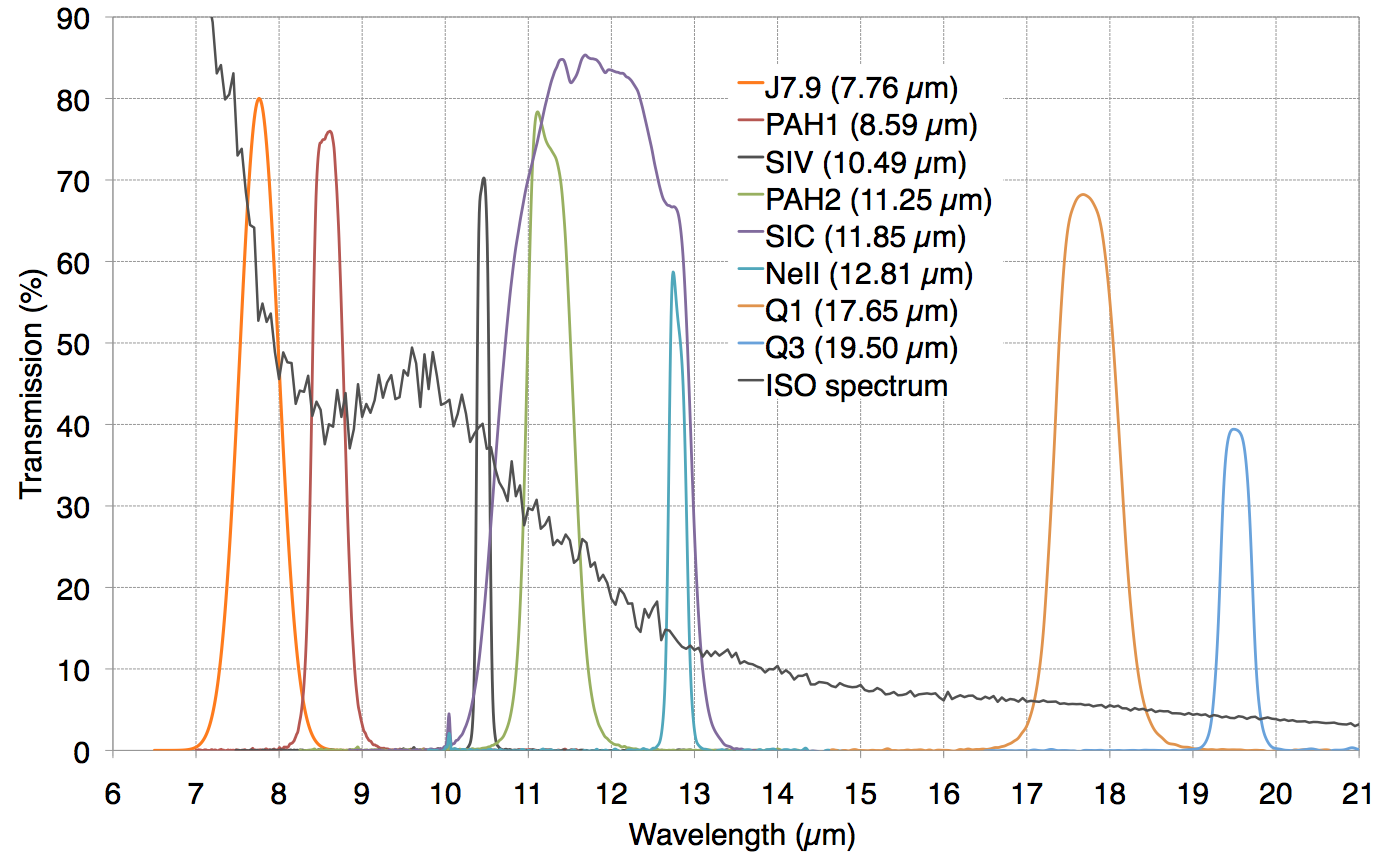}
\caption{Transmission of the VISIR filters used for the observations of Betelgeuse. The transmission of the J7.9 filter is supposed to be Gaussian with a central wavelength of 7.76\,$\mu$m and a half-band width of 0.55\,$\mu$m. The ISO spectrum of Betelgeuse is also shown for reference, with an arbitrary intensity scale (thin grey curve, see Sect.~\ref{OverallNebula} for details). The central wavelength of each filter is given in the legend.\label{VISIR_filters}}
\end{figure}

We observed Betelgeuse and Aldebaran on 12 November 2010, between UT 06:54 and 08:04 (Table~\ref{visir_log}). We recorded BURST image cubes of several thousand images in eight VISIR filters: J7.9 ($\lambda_0=7.76\,\mu$m, half-band width $\Delta\lambda=0.55\,\mu$m), PAH1 ($\lambda_0=8.59 \pm 0.42$\,$\mu$m), SIV ($\lambda_0=10.49\,\mu$m, $\Delta\lambda=0.16\,\mu$m), PAH2 ($\lambda_0=11.25\,\mu$m, $\Delta\lambda=0.59\,\mu$m), SIC ($\lambda_0=11.85\,\mu$m, $\Delta\lambda=2.34\,\mu$m), NeII ($\lambda_0=12.81\,\mu$m, $\Delta\lambda=0.21\,\mu$m), Q1 ($\lambda_0=17.65\,\mu$m, $\Delta\lambda=0.83\,\mu$m), and Q3 ($\lambda_0=19.50\,\mu$m, $\Delta\lambda=0.40\,\mu$m). Their transmission curves are shown in Fig.~\ref{VISIR_filters}, and are available from the VISIR instrument web site at ESO\footnote{http://www.eso.org/sci/facilities/paranal/instruments/visir/}.
The images were chopped and nodded in a North-South and East-West direction by offsetting the M2 mirror and the telescope (respectively) in order to remove the fluctuating thermal background in the post-processing. The chopping and nodding amplitudes were both set to 8" on the sky. The chopping frequency was set to 0.5\,Hz in the $N$ band and 1\,Hz in the $Q$ band, and the nodding period was set to 90\,s.

\subsection{Raw data processing}

The data processing procedure we developed to reduce the data cubes is described in Kervella \& Domiciano de Souza~(\cite{kervella07}). After a classical chopping-nodding subtraction of the background, the frames were precisely recentered. We then spatially resampled each individual frame by a factor four. This oversampling allows us to take advantage of the atmospheric jitter of the images to improve the sampling of the point spread function (through a classical dithering technique). To compute the final images, we averaged the complete image cubes in all filters, as the improvement in angular resolution brought by image selection was marginal, while degrading the sensitivity to the faint extensions of the nebula.

\subsection{Treatment of image saturation\label{saturation}}

The extreme brightness of our two target stars resulted in a saturation of the VISIR detector on the PSF cores in some of the filters. For Aldebaran, some limited saturation occured with the PAH1, SIV, and PAH2 filters. For Betelgeuse, all filters in the $N$ band showed some degree of saturation: low for J7.9 and NeII, moderate for SIC, and heavy for PAH1, SIV and PAH2. The Q1 and Q3 filters did not produce saturation for the two stars. The effect of saturation of the VISIR images is two-fold: firstly, it causes an underestimation of the total photometry of the saturated image, and secondly, it creates a number of image artefacts.

\subsubsection{PSF core saturation}

\begin{figure}
\centering
\includegraphics[width=\hsize]{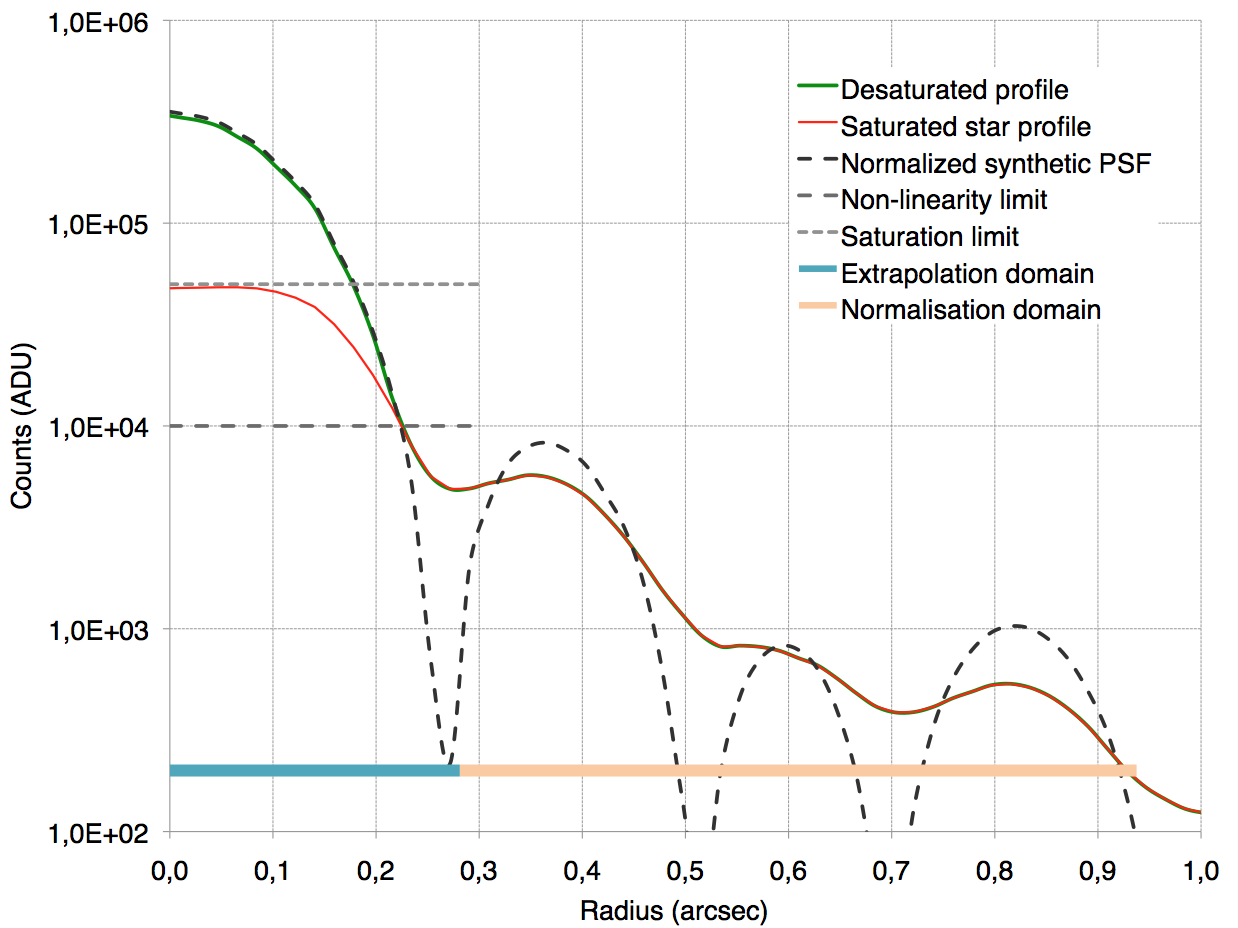}
\caption{Principle of the restoration of the saturated PSF core of the images of Aldebaran. The curves represent radial median profiles of the original saturated image (thin red curve), the desaturated image (thick green curve) and the synthetic PSF used for the central extrapolation (dashed grey curve). The non-linearity and saturation limits are represented with dashed and dotted grey lines. 
\label{desaturation_PAH1}}
\end{figure}

The effect of the photometric bias due to saturation is different for Betelgeuse and Aldebaran. For Betelgeuse, the bias on the photometry of the PSF core is not a problem since we are interested in the CSE, that, although rather bright, does not saturate the detector. For Aldebaran (PSF and photometric calibrator), the saturation of the PSF core in the PAH1, SIV and PAH2 filters creates a bias on the total flux, and makes the images unusable for the deconvolution of the images of Betelgeuse. To overcome this problem, we extrapolated the saturated PSF core of Aldebaran using a synthetic PSF profile. It was computed taking into account the shape of the telescope pupil (including the secondary mirror obscuration), the photometric transmission of the atmosphere (with airmass dependence), the transmission of the VISIR filters, and the quantum efficiency of the detector. The resulting synthetic PSF images were normalized to the observed flux in the images of Aldebaran, using as normalization a ring centered on the star and avoiding the saturated area. In practice, we integrated the flux over a ring from $\approx 0.3\arcsec$ to $\approx 0.9\arcsec$ from the star. The result of the PSF core extrapolation, as well as the normalization and integration domains are shown in Fig.~\ref{desaturation_PAH1}. The normalized synthetic PSF core was used for the extrapolation only where its value exceeded that of the original saturated image. This resulted in a continuous and smooth transition between the synthetic extrapolated core and the external part of the PSF.

\subsubsection{Cosmetic artefacts}

\begin{figure}
\centering
\includegraphics[width=4.4cm]{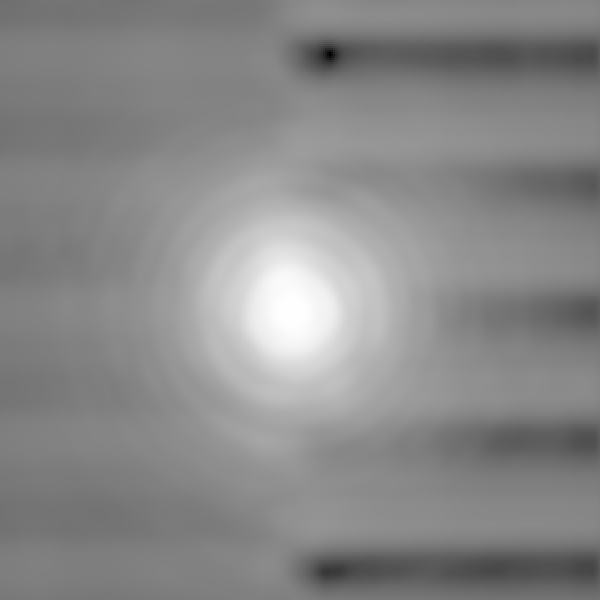}
\includegraphics[width=4.4cm]{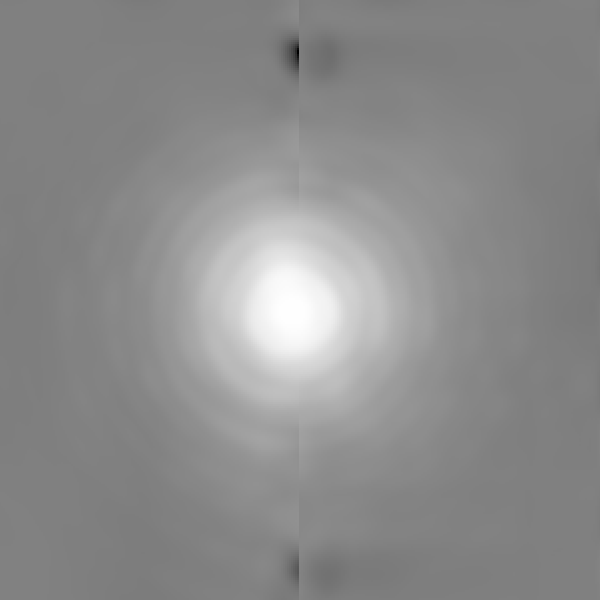}
\caption{Stripping of the Betelgeuse average image in the PAH1 filter caused by heavy saturation (left panel) and corrected version (right). The scale is logarithmic to enhance the faint extensions of the image.
\label{dirty_bet_PAH1}}
\end{figure}

\begin{figure*}[]
\centering
\includegraphics[width=4.5cm]{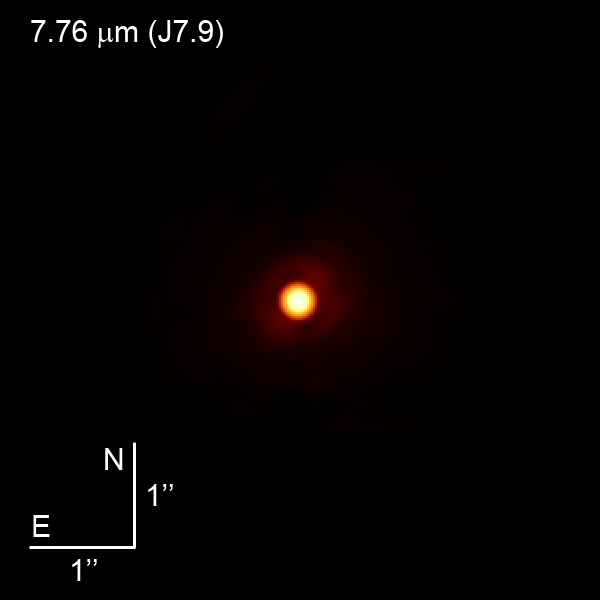} \includegraphics[width=4.5cm]{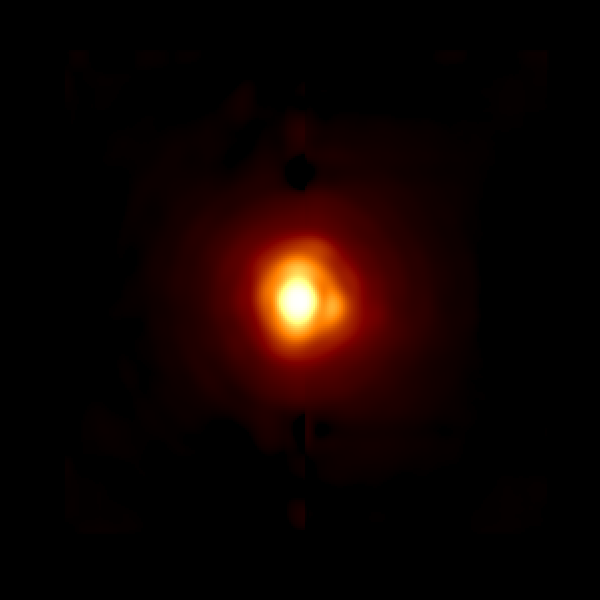} \hspace{1mm}
\includegraphics[width=4.5cm]{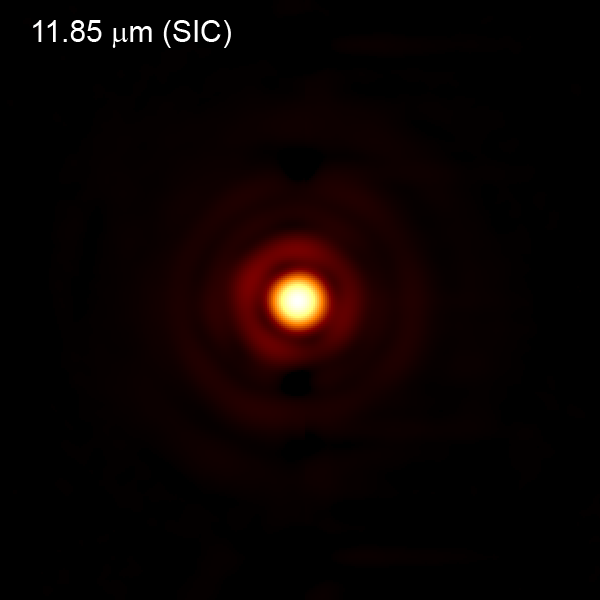} \includegraphics[width=4.5cm]{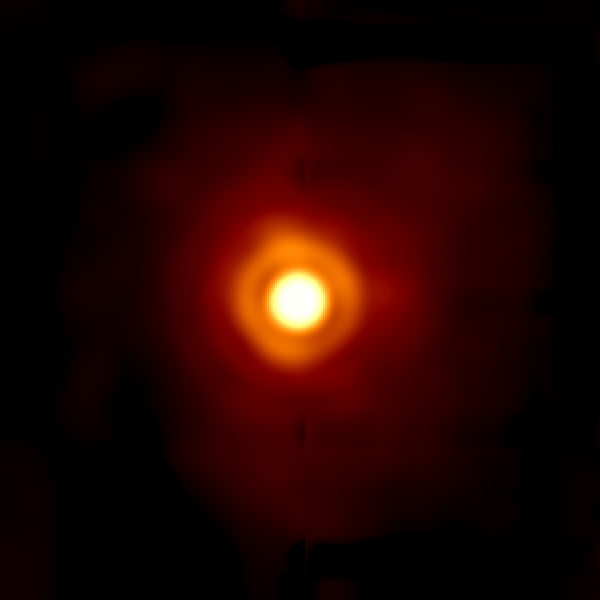}

\includegraphics[width=4.5cm]{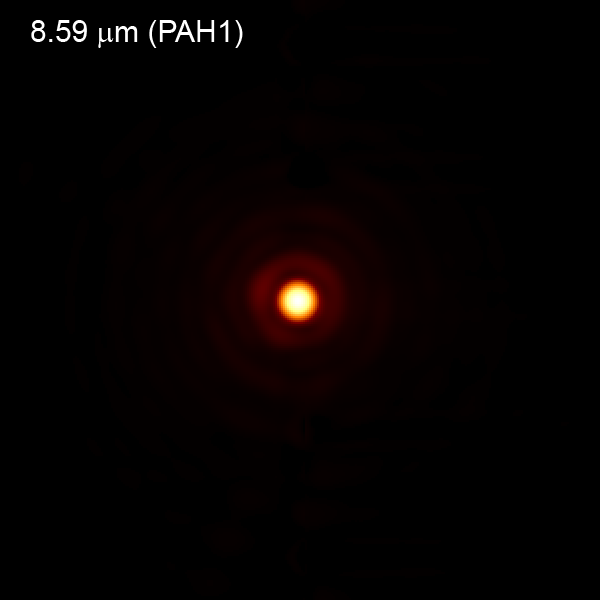} \includegraphics[width=4.5cm]{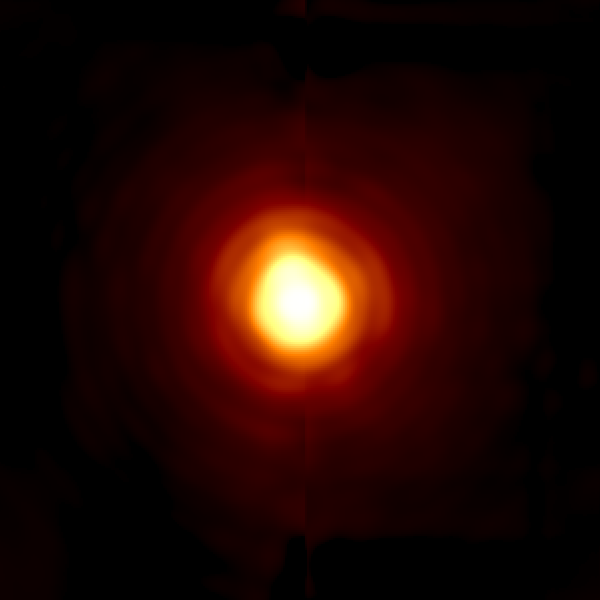} \hspace{1mm}
\includegraphics[width=4.5cm]{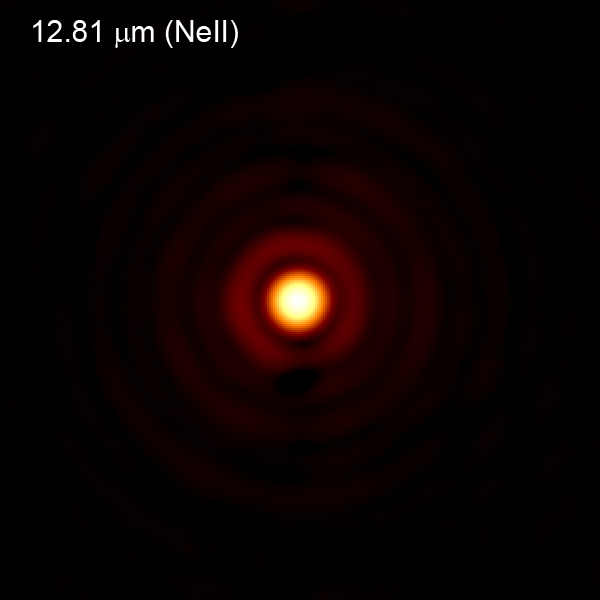} \includegraphics[width=4.5cm]{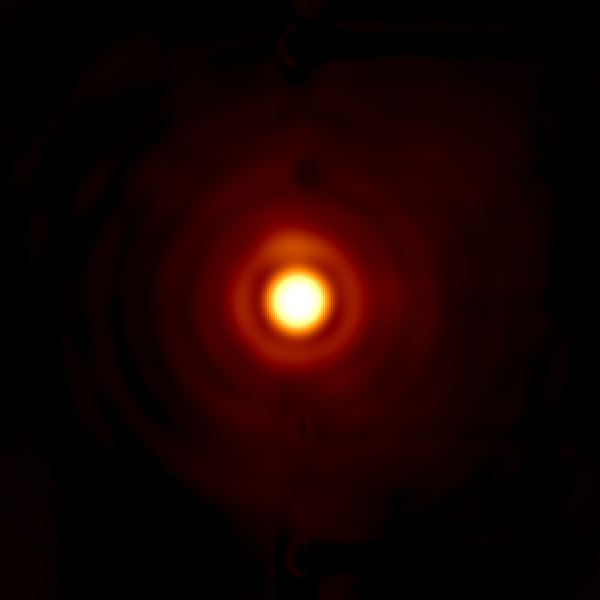}

\includegraphics[width=4.5cm]{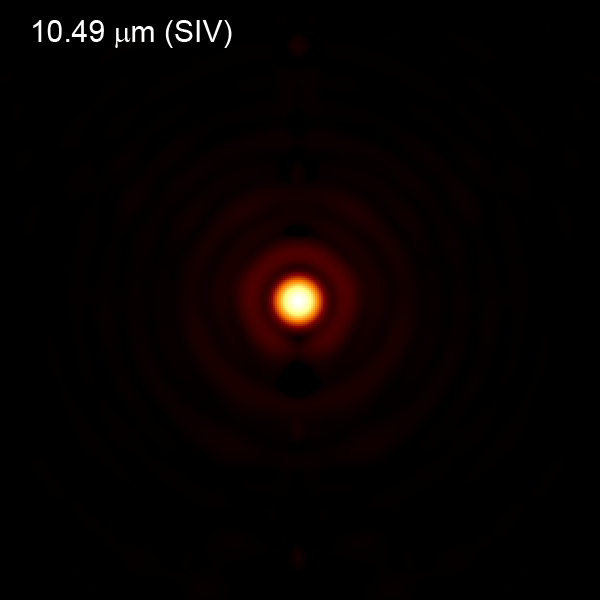} \includegraphics[width=4.5cm]{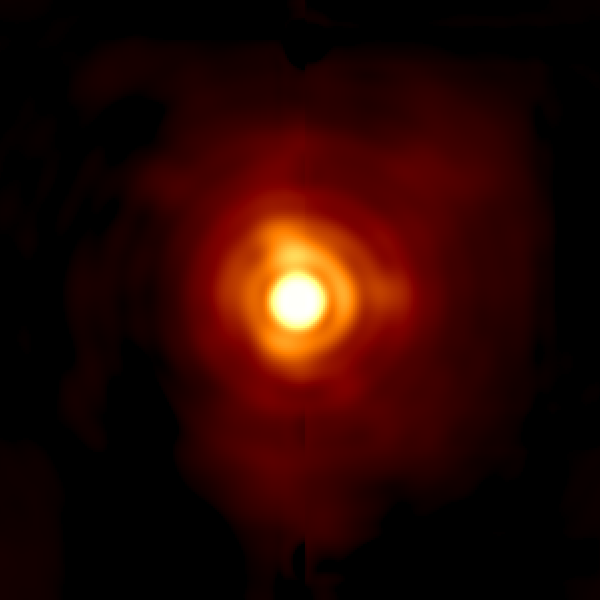} \hspace{1mm}
\includegraphics[width=4.5cm]{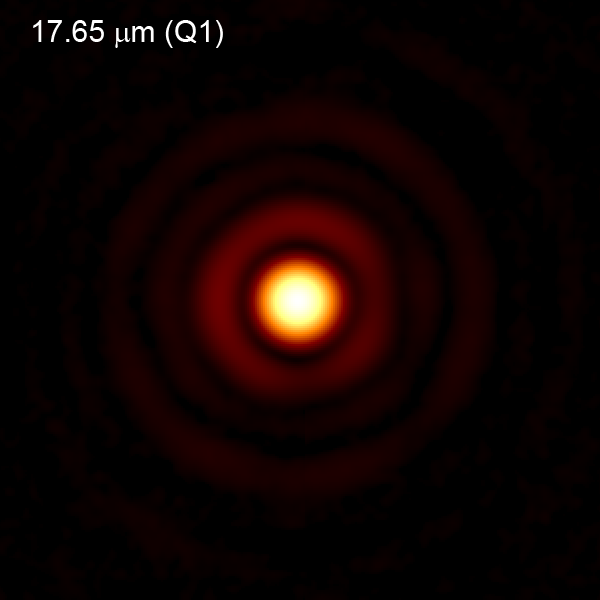} \includegraphics[width=4.5cm]{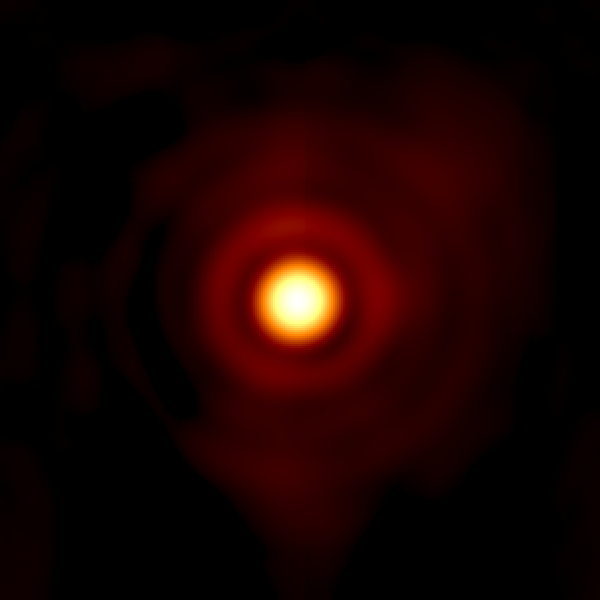}

\includegraphics[width=4.5cm]{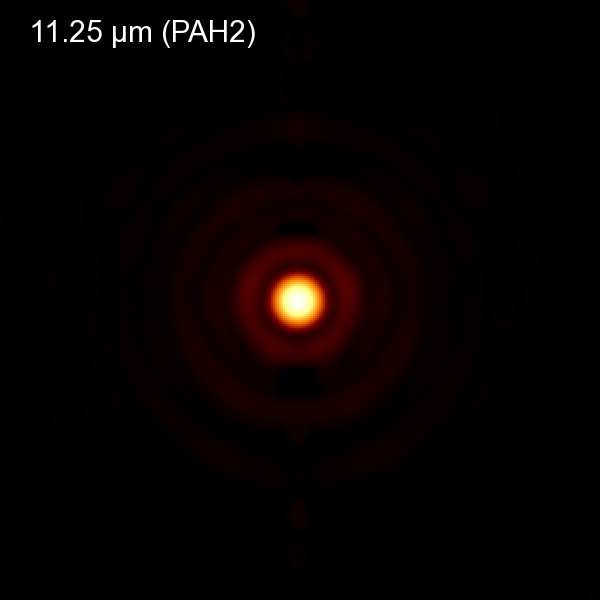} \includegraphics[width=4.5cm]{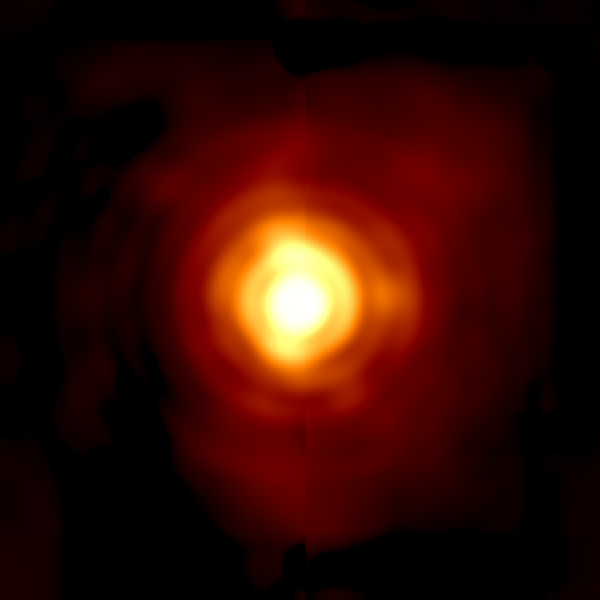} \hspace{1mm}
\includegraphics[width=4.5cm]{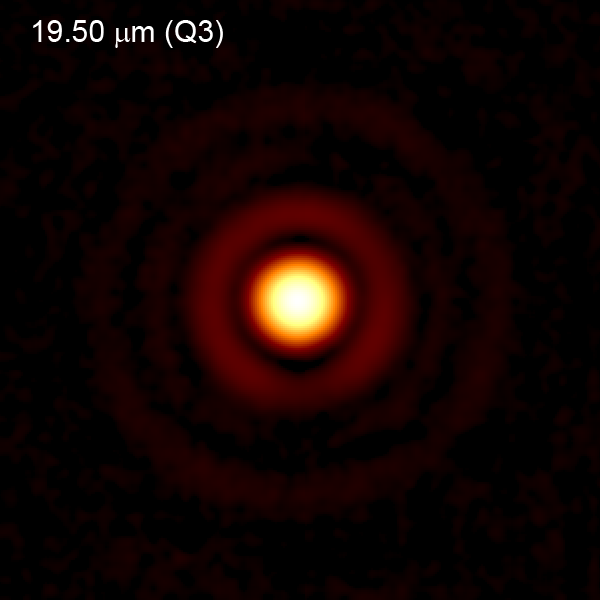} \includegraphics[width=4.5cm]{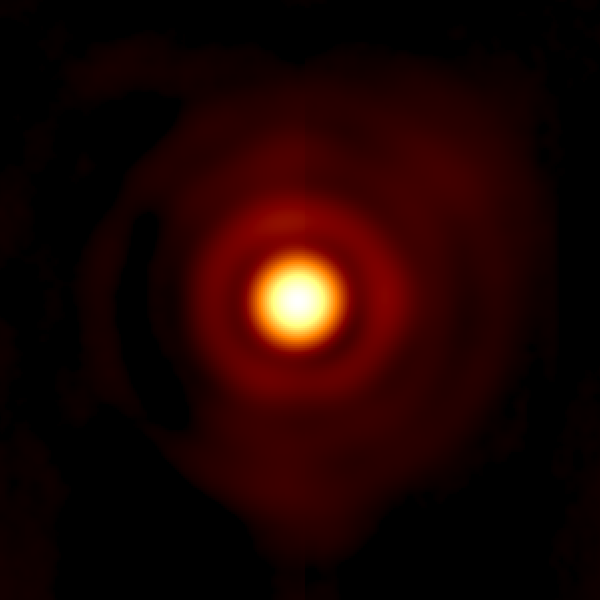}

\caption{Average VISIR images of the PSF calibrator Aldebaran (left image of each column) and Betelgeuse (right image of each column) in 8 filters from 7.76 to 11.25\,$\mu$m (left column) and 11.85 to 19.50\,$\mu$m (right column). The color scale is a function of the square root of the intensity, scaled to the minimum and maximum values in each image. The field of view is $5.63\arcsec \times 5.63\arcsec$ for all images, with North up and East to the left. The images of Aldebaran in the PAH1, SIV and PAH2 filters, as well as all the $N$ band images of Betelgeuse include a correction of the PSF core saturation (see Sect.~\ref{saturation} for details). \label{Avg_cubes}}
\end{figure*}

Another effect of saturation is the presence of several types of cosmetic artefacts, mainly stripes, on the images. The memory effect of the detector also causes some ghosts to appear at the chopping positions where the star is positioned, but this only results in a moderate photometric bias, on the saturated part of the PSF, that is unimportant for our analysis. The striping effect is more problematic, as it creates a noticeable pattern over the image, that directly affects the photometry of the nebula. Fortunately, the removal of this pattern is efficiently obtained by subtracting the median value of each line computed over a segment near the right and left edges of the frame (Fig.~\ref{dirty_bet_PAH1}). Some saturation-related artefacts are still present close to the star, in particular directly above and below, but they remain at an acceptable level in all filters. The resulting images of Betelgeuse and Aldebaran in all filters are presented in Fig.~\ref{Avg_cubes}.

\subsection{PSF subtraction and photometric calibration}

The CSE of Betelgeuse is contaminated by the numerous diffraction rings from the (unresolved) central star. These rings present a significant intensity compared to the nebula, and even dominate close to the star. It is therefore important to subtract them properly. The core of the PSF of Betelgeuse being saturated in all $N$-band filters, the normalization of the PSF images is particularly difficult. It is also not possible to use the unsaturated part of the images of Betelgeuse (e.g. the external diffraction rings) to normalize the PSF image, as the nebula contributes significantly to the total flux and would cause an overestimation of the normalization factor. Due to these constraints, we adopted an original approach based on theoretical models of the spectral energy distribution of Betelgeuse (the star itself, excluding the envelope) and Aldebaran.

\begin{figure}
\centering
\includegraphics[width=4.4cm]{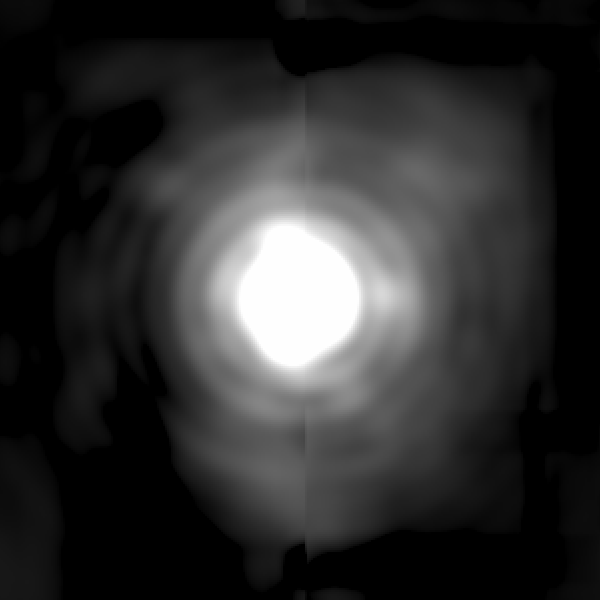}
\includegraphics[width=4.4cm]{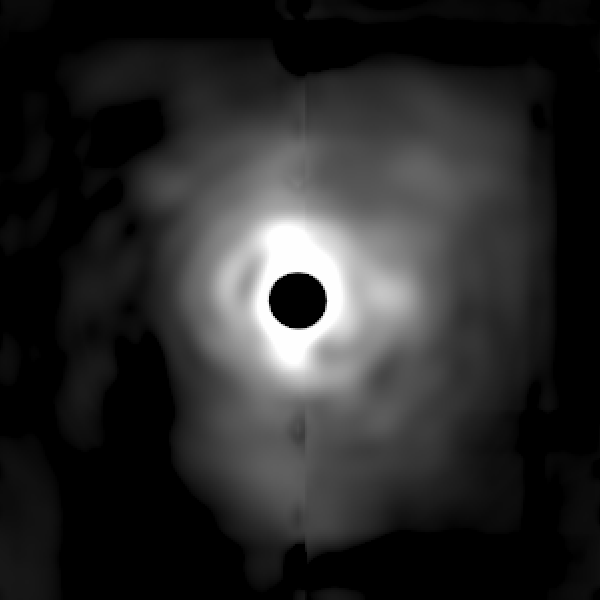}
\caption{Image of Betelgeuse in the PAH2 filter before (left) and after (right) the subtraction of the PSF image derived from the observation of Aldebaran. The grey scale is a function of the square root of the intensity between 0 and 0.5\,W\,m$^{-2}$\,$\mu$m$^{-1}$\,sr$^{-1}$. The black disk in the center corresponds to the non-linear and saturated part of the Betelgeuse image that was masked after the subtraction.
\label{bet_psfsub}}
\end{figure}

We used the broadband magnitudes of Betelgeuse and Aldebaran from the literature to calibrate Castelli \& Kurucz~(\cite{castelli03}) models of their spectral energy distributions (see Paper~I for details). We used these models to calibrate photometrically the images of Aldebaran, and derive the ratio of the apparent fluxes of the two stars in all our filters (including the atmosphere and detector transmissions). We then subtracted the scaled images of Aldebaran from the Betelgeuse images. Using this procedure, we subtracted the photometric contribution from the central star, but we leave the envelope untouched. A comparison of an image of Betelgeuse before and after the PSF subtraction is presented in Fig.~\ref{bet_psfsub}. The numerous diffraction rings of Betelgeuse are efficiently subtracted and do not appear on the resulting envelope image. The main geometrical characteristics of the shell can clearly not be attributed to the PSF subtraction as the PSF is circularly symmetric and as the departure from symmetry of the shell is visible on the raw images.

\subsection{Image deconvolution}

\begin{figure*}[]
\centering
\includegraphics[width=4.5cm]{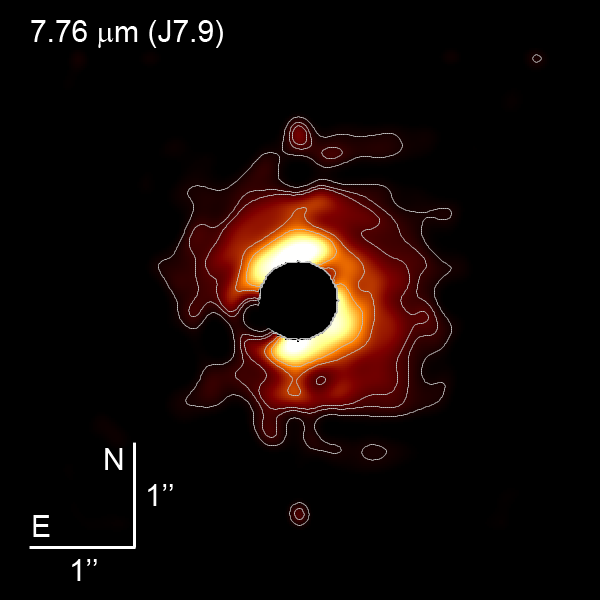} \includegraphics[width=4.5cm]{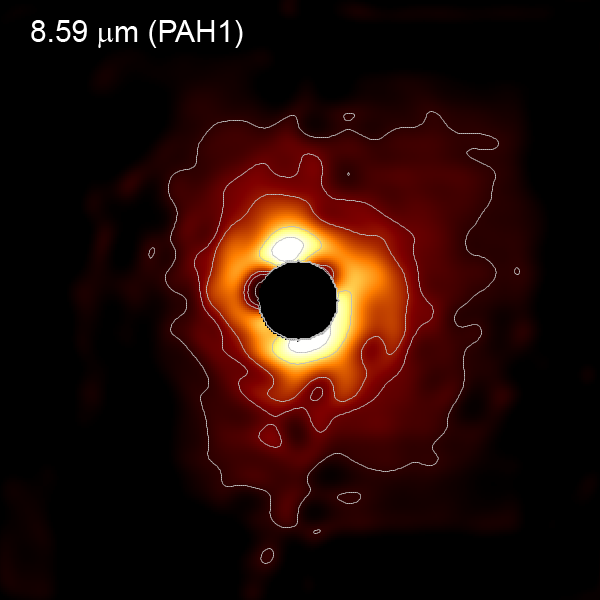}
\includegraphics[width=4.5cm]{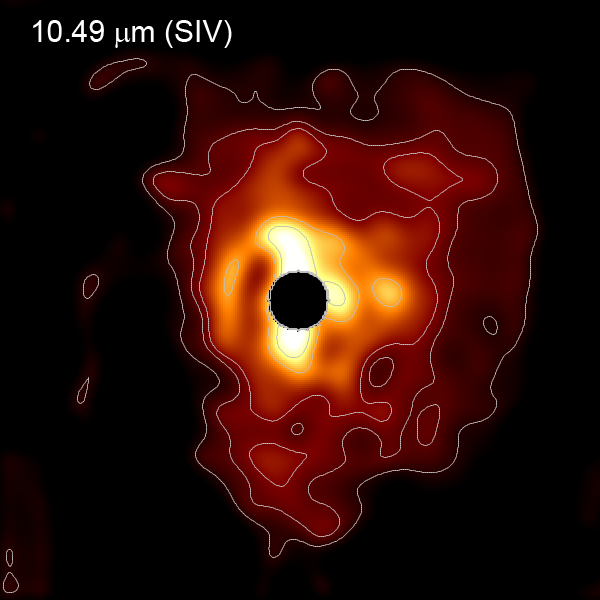} \includegraphics[width=4.5cm]{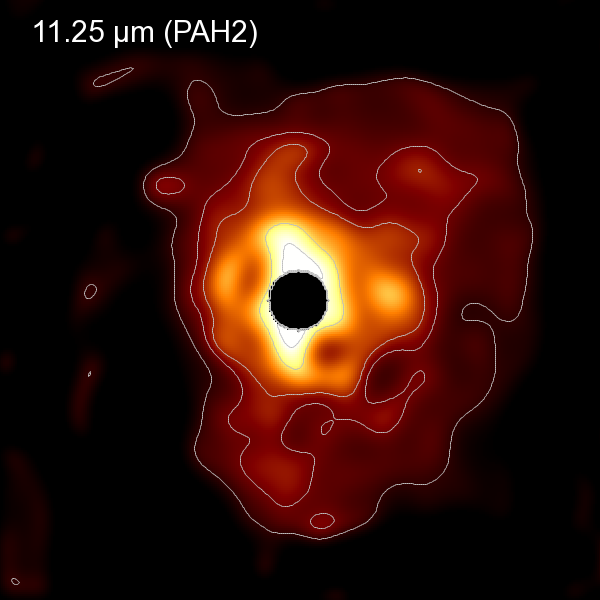} \\
\includegraphics[width=4.5cm]{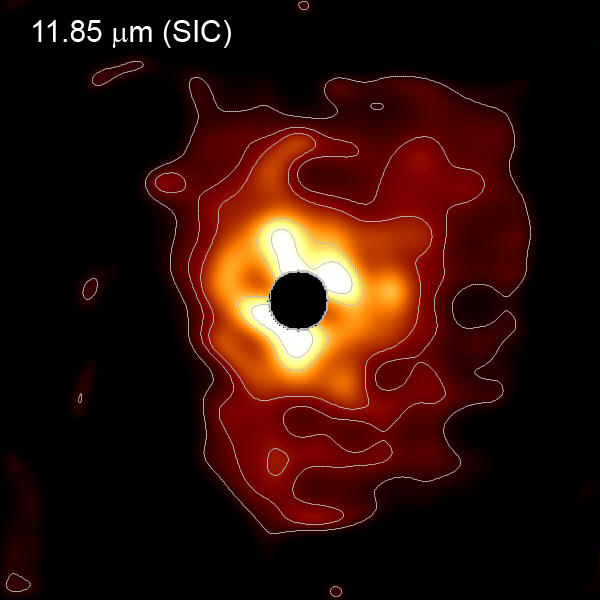} \includegraphics[width=4.5cm]{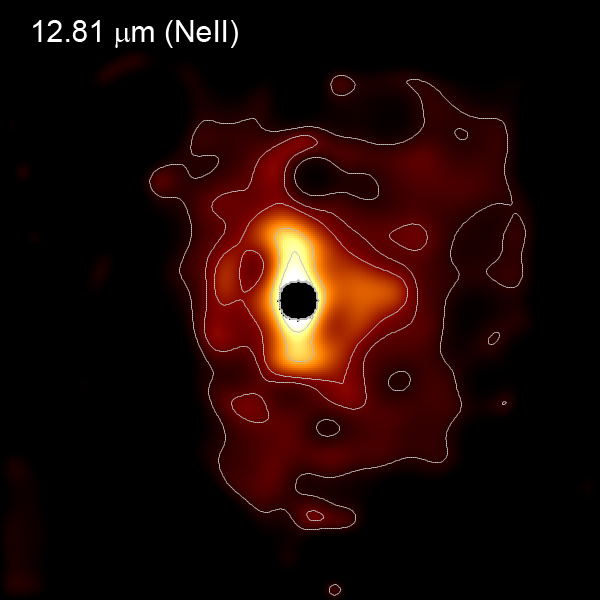} 
\includegraphics[width=4.5cm]{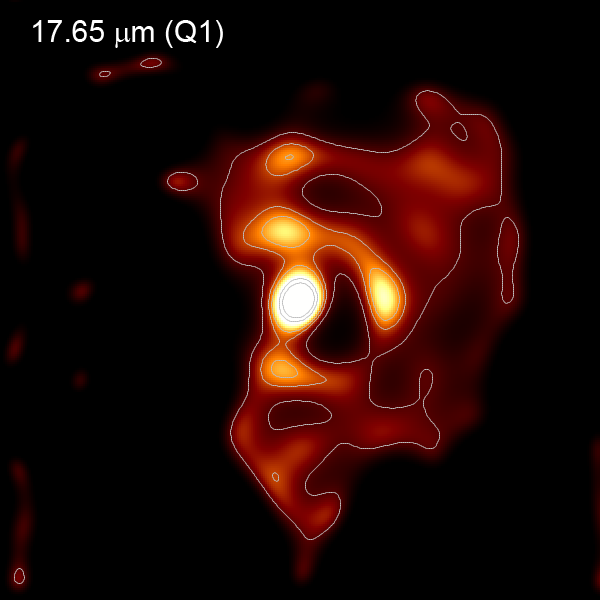} \includegraphics[width=4.5cm]{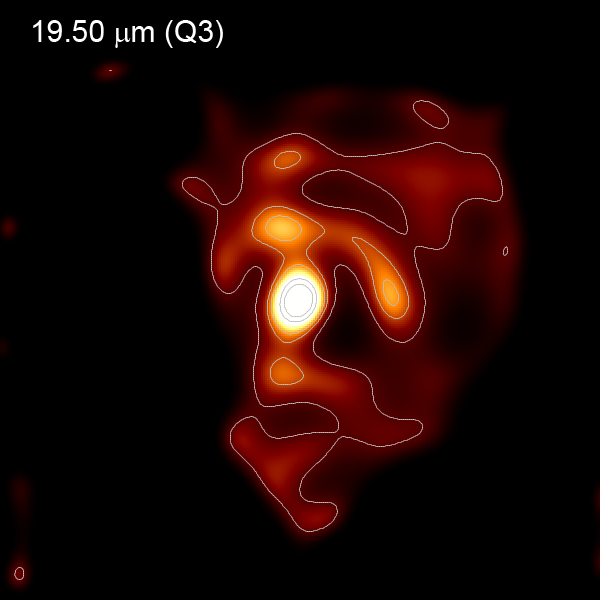} \\
\caption{PSF subtracted and deconvolved images of the CSE of Betelgeuse for the eight VISIR filters in the $N$ and $Q$ bands. The field of view is $5.63\arcsec \times 5.63\arcsec$ for all images, with North up and East to the left. The color scale is a function of the square root of the intensity and is normalized to the maximum and minimum intensity in the image. The contour levels correspond to 0.01, 0.05, 0.1, 0.5 and 1.0 W\,m$^{-2}$\,$\mu$m$^{-1}$\,sr$^{-1}$. For the $N$ band images, the dark spot close to the center corresponds to the saturated part of the image that was ignored in the deconvolution process.\label{PSF_decLR_subNQ}}
\end{figure*}

The VISIR images obtained using the BURST mode are well suited for deconvolution, as they are essentially diffraction limited, and mostly insensitive to the seeing.
We deconvolved the PSF subtracted images of Betelgeuse using the images of Aldebaran (corrected for core saturation in the PAH1, SIV and PAH2 filters) as the dirty beams, and IRAF\footnote{http://iraf.noao.edu/}'s implementation of the Lucy-Richardson (L-R) algorithm ({\tt lucy} command of the {\tt stsdas}\footnote{http://www.stsci.edu/resources/software\_hardware/stsdas} package).
We stopped the L-R deconvolution after a number of iterations depending on the wavelength: 5 iterations for J7.9 and PAH1, 10 iterations for SIV and PAH2, 20 iterations for SIC and NeII, and 40 iterations for the Q1 and Q3 filters. This variable number of iterations results in an homogeneous deconvolved image quality for all filters, while avoiding the creation of deconvolution artefacts. This is particularly important at the shortest wavelengths (J7.9 and PAH1 filters), where the nebula is faint and the PSF subtraction is delicate. In the $Q$ band, there is no saturation, and the PSF subtraction is much easier and efficient. This allowed us to use a larger number of deconvolution steps.
The resulting deconvolved images of the nebula of Betelgeuse in the six $N$ band and the two $Q$ band filters are presented in Fig.~\ref{PSF_decLR_subNQ}.

\section{Image analysis \label{analysis}}

\subsection{Structure of the nebula}

\begin{figure}[]
\centering
\setlength\fboxsep{0pt} 
\setlength\fboxrule{0.5pt}
\fbox{\includegraphics[width=\hsize]{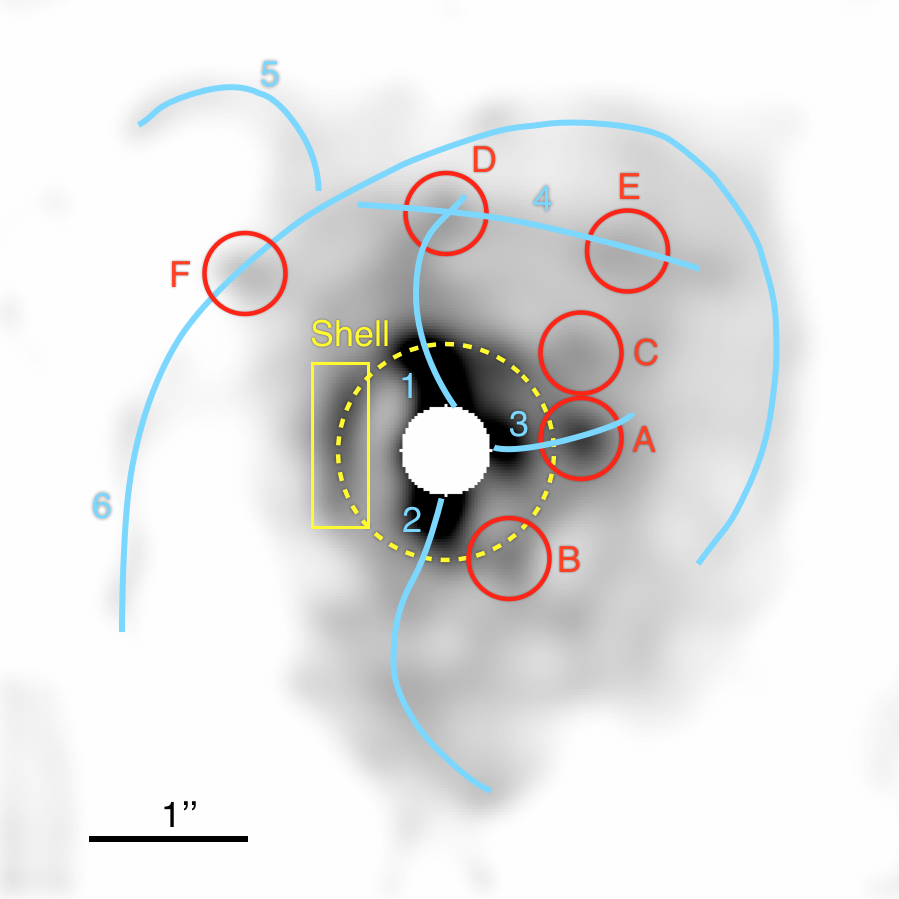}}
\caption{Main identified features in the circumstellar nebula of Betelgeuse. The background is the image in the SIV filter, with an inverted grey scale proportional to the square root of the flux. \label{Features}}
\end{figure}

The overall atructure of the nebula appears to be approximately triangular, with an inhomogeneous surface brightness. Extensions are present up to $\approx 3\arcsec$ from the star to the North, South and Northwest (Fig.~\ref{Features}). An irregular ring-like structure is visible at a radius of $0.5$ to $1.0\arcsec$, particularly in the Eastern part of the SIV filter image (rectangle in Fig.~\ref{Features}).
For concision, we will refer to this structure as ``the shell", although we do not have evidence from our images of its actual tridimensional shape.
As shown in Fig.~\ref{PSF_decLR_subNQ}, the shell is also detectable in the other $N$ band filters. In the two $Q$ band filters, it is clearly visible in the north-western quadrant of the image, at a slightly larger angular radius from the star.

The plumes labeled 1, 2 and 3 appear to brighten inside the shell, and may originate from the star itself. Several bright knots are also visible in the nebula (labels A to F, ordered by decreasing brightness in the SIV filter). Some of these knots may be connected to the star through the plumes (knots A and C). Additional filamentary structures are observed at distances of 2 to $3\arcsec$ from the star (labels 4 and 5), and a large elliptical loop is marginally detectable in the SIV, PAH2 and SIC images. The coordinates of the photometric measurement windows over the knots and the shell are listed in Table~\ref{feature_coords}.

\begin{table}
\caption{Coordinates of the photometric measurement areas shown in Fig.~\ref{Features} in the nebula of Betelgeuse, relatively to the star. } 
\label{feature_coords}
\begin{tabular}{lrrrr}
\hline \hline
Feature & $\Delta \alpha$ & $\Delta \delta$ & Radius & PA \\
 & $(\arcsec)$ & $(\arcsec)$ & ($\arcsec$) & ($^\circ$) \\
\hline
\noalign{\smallskip}
A & $-0.88$ & 0.04 & 0.88 & 272 \\
B & $-0.39$ & $-0.71$ & 0.81 & 209 \\
C & $-0.84$ & 0.62 & 1.05 & 306 \\
D & 0.00 & 1.48 & 1.48 & 0 \\
E & $-1.11$ & 1.24 & 1.66 & 318 \\
F & 1.24 & 1.09 & 1.65 & 49 \\
\hline
\noalign{\smallskip}
Shell arc center & 0.66 & 0.04 & 0.66 & 87 \\
Shell SE point & 0.84 & $-0.47$ & 0.97 & 119 \\
Shell NW point & 0.47 & 0.54 & 0.72 & 41 \\
\noalign{\smallskip}
\hline
\end{tabular}
\tablefoot{The position angle (PA) is counted positively from North to East (N=$0^\circ$, E=$90^\circ$). The last three lines list the positions of the center of the shell measurement window, and its south-eastern and north-western corners.}
\end{table}

\subsection{Overall spectral energy distribution of the nebula \label{OverallNebula}}

We derived the spectral energy distribution of the nebula in our eight VISIR filters through an integration of the flux over the PSF-subtracted , but non-deconvolved images. We did not use the deconvolved images, in order to avoid possible biases due to the deconvolution algorithm, and also because we did not need the high angular resolution. We integrated the flux of the nebula over a circular aperture of $6\arcsec$ in diameter. The central part of the nebula is lost during the PSF subtraction, due to the saturation of the core of the images of Betelgeuse (except in the $Q$ band). To avoid a bias on the integrated photometry, we extrapolated the nebular flux in the central masked disk ($0.4\arcsec$ in radius) using a Gaussian fit over the rest of the PSF-subtracted image of the envelope. This resulted in practice in a marginal additional contribution for all filters ($<3\%$), except in the J8.9 and PAH1 filters ($\approx 20\%$ and $10\%$, respectively).

\begin{figure}[]
\centering
\includegraphics[width=\hsize]{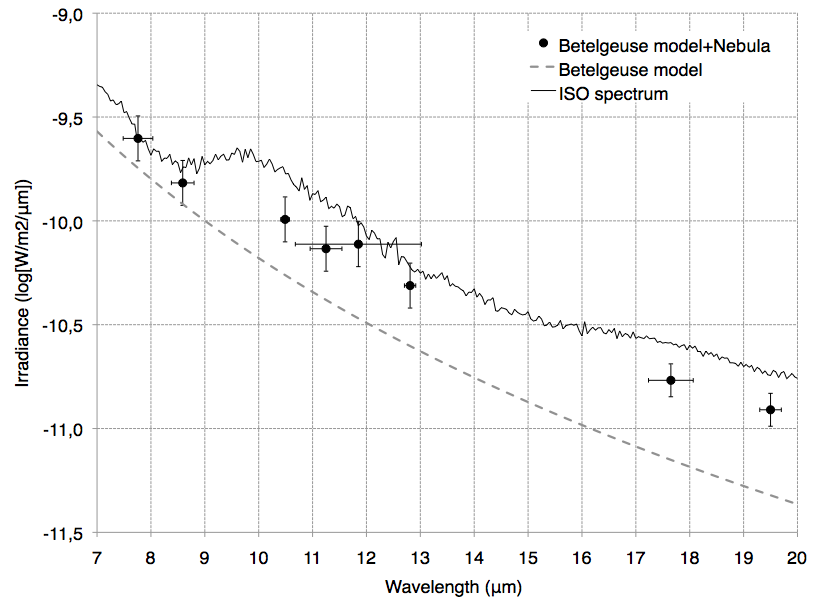}
\caption{Integrated irradiance of Betelgeuse plus the nebula (black dots), superimposed on the model spectrum of the star alone (dashed curve), compared to the ISO SWS spectrum of Betelgeuse. The horizontal segments attached to each point indicate the spectral width at half transmission of the VISIR filters.\label{TotalFlux}}
\end{figure}

The estimation of the uncertainty in the determination of the absolute flux of Betelgeuse is difficult, due to the treatment of saturation that we had to apply to the PSF reference images. Thanks to the high brightness of the two stars, the statistical dispersion of the photometry is negligible, compared to the systematics. We estimate that the systematic uncertainty on our measurement of the irradiance of Aldebaran is $\approx 20\%$, and that it directly applies to the photometry of the nebula of Betelgeuse. In Fig.~\ref{TotalFlux}, we show the total irradiance of the circumstellar nebula as a function of wavelength. The systematic uncertainty on our model of the irradiance of Betelgeuse itself (the star alone, for the PSF subtraction) is estimated also to 20\%, due essentially to the variability of the star. Overall, this results in $\approx 30$\% error bars for the $N$ band measurements (Betelgeuse model uncertainty plus saturation correction uncertainty), and $\approx 20\%$ in the $Q$ band (Betelgeuse model uncertainty alone).

Fig.~\ref{TotalFlux} also shows for comparison the spectrum of Betelgeuse obtained using ISO's (Kessler et~al.~Ê\cite{kessler96}) SWS instrument (de Graauw et~al.~\cite{degraauw96}), retrieved from the ISO Data Archive\footnote{http://iso.esac.esa.int/ida/, observation ID 69201980}. The overall shape of the spectrum is comparable to our VISIR photometric data points, although slightly brighter. This systematic difference at a $1-2\sigma$ level could be explained by the relatively large aperture diameter of the SWS spectrograph (14 to $27\arcsec$, vs $6\arcsec$ for our flux integration disk), or by the temporal variability of Betelgeuse (the ISO spectrum was obtained on 1997-10-08). The absolute photometric uncertainty of SWS low-resolution spectra is approximately 5-10\%, and may also explain part of the observed difference.

\subsection{Spectral energy distribution of individual features}

We measured the surface brightness of the different identified features of the nebula on the PSF-subtracted and deconvolved images of Betelgeuse. We prefered the deconvolved images for this task for two reasons. Firstly, McNeil \& Moody~(\cite{mcneil05}) showed that the Lucy-Richardson algorithm preserves the photometric accuracy relatively well compared to other classical deconvolution algorithms. Secondly, our images cover a very broad range of wavelengths, with more than a factor 2.5 between our extreme wavelengths (J7.9 and Q3 filters). The diffraction-limited angular resolution therefore also varies by this factor, and the surface photometry over a given aperture size would result in very different actual measurement areas on the sky depending on the filter (the adjacent regions to the aperture would contribute differently depending on the angular resolution). For these reasons, we prefered to use the deconvolved images for the photometry presented in this Section.
%
\begin{figure}[h]
\centering
\includegraphics[width=\hsize]{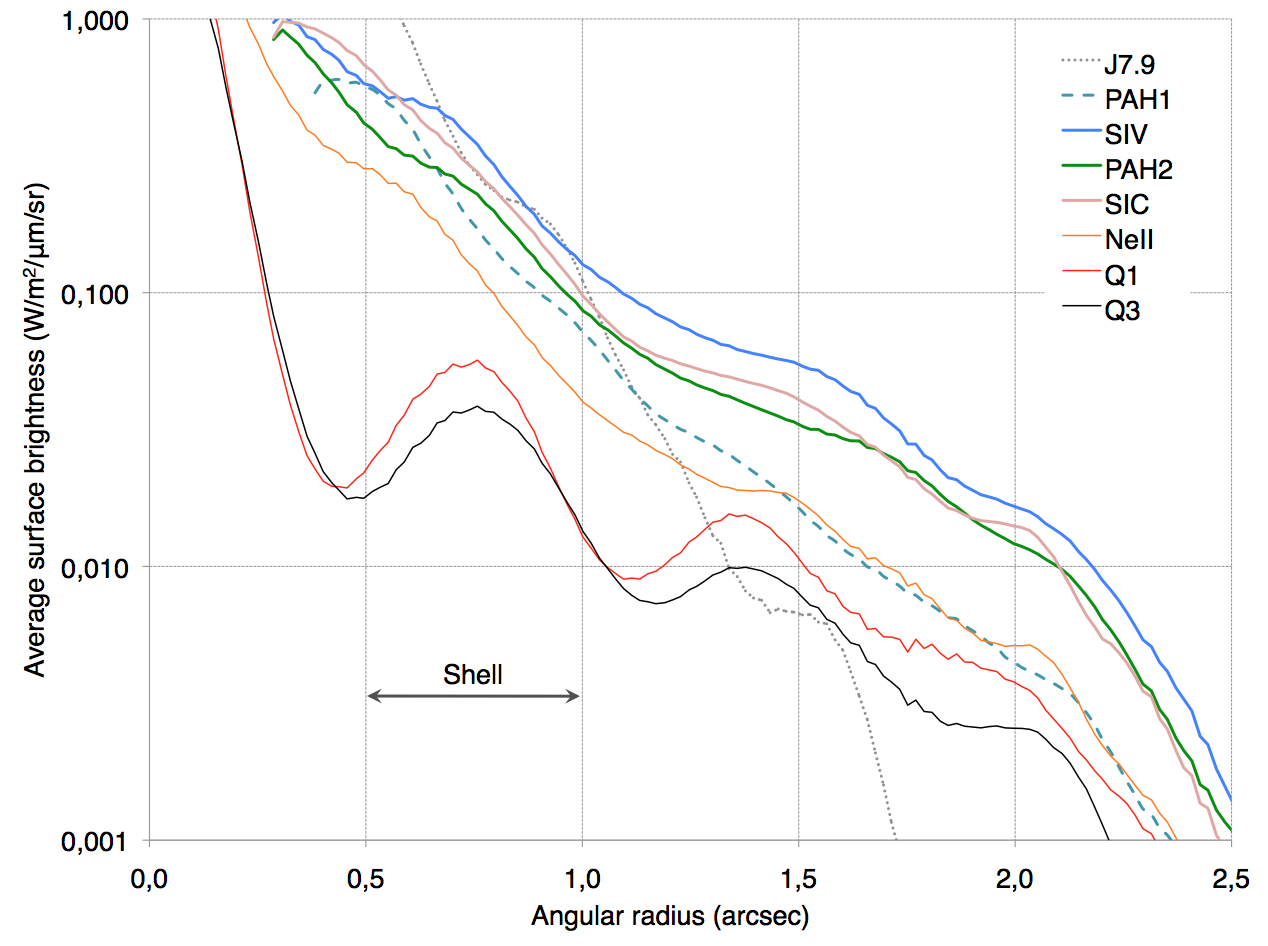}
\caption{Azimuthally-averaged surface brightness profile of the CSE of Betelgeuse, computed from the PSF-subtracted and deconvolved images.\label{radprof}}
\end{figure}
The azimuthally-averaged surface brightness of the CSE of Betelgeuse is represented in Fig.~\ref{radprof} as a function of the angular separation from the star. The approximate radius of the shell is shown for convenience.

\subsubsection{The shell}

\begin{figure}[]
\centering
\includegraphics[width=\hsize]{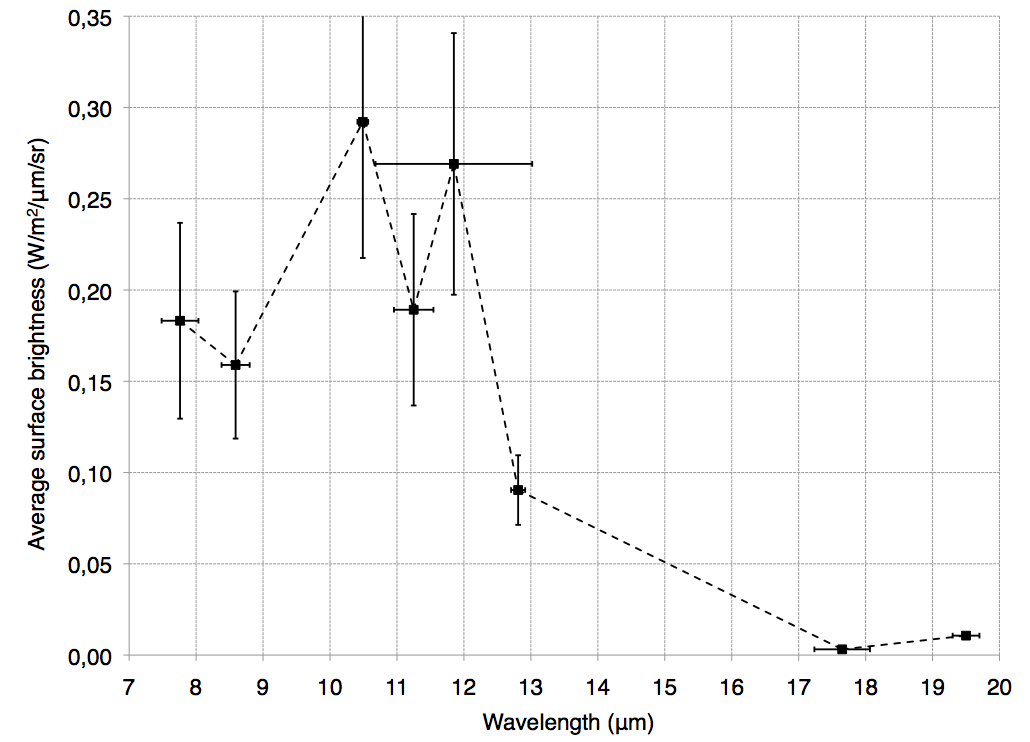}
\caption{Average surface brightness over a rectangular area located on the Eastern section of the shell (defined in Fig.~\ref{Features}). The vertical error bars correspond to the standard deviation of the surface brightness over the measurement area, and the horizontal error bars indicate the spectral width at half transmission of the VISIR filters. \label{Shell}}
\end{figure}

The approximate location of this ring-like structure, that we tentatively interpret as a shell, is represented with a dashed circle in Fig.~\ref{Features}. Although its overall shape is circular, it is relatively irregular in its western section, whereas its eastern part shows a mostly continuous arc in the $N$ band. In the $Q$ band, the western section is more clearly defined (Fig.~\ref{PSF_decLR_subNQ}). We nevertheless chose the eastern section of the shell to estimate its surface brightness (rectangular aperture drawn in Fig.~\ref{Features}), as it is visible in all $N$ band filters. The coordinates of the corners and center of the measurement window are listed in Table~\ref{feature_coords}. The resulting photometry is presented in Fig.~\ref{Shell}. The shell appears particularly bright in the SIV, PAH2 and SiC filters which is not systematically the case closer to the star. Another interesting property is that it appears brighter in the Q3 band ($10.7 \pm 1.1$~mW/m$ ^2$/$\mu$m/sr) than in the Q1 band ($3.2 \pm 1.2$~mW/m$ ^2$/$\mu$m/sr). This spectro-photometric pattern is compatible with the presence of typical O-rich dust, such as silicate-type materials.
In the average radial profile of the CSE shown in Fig.~\ref{radprof}, the photometric contribution of the shell is visible in the SIV, PAH2 and SIC filters as a ``bump" in the azimuthaly-averaged profile. It is also easily observable in the $Q$ band filters, thanks to its higher contrast compared to the nebular background.

\subsubsection{Nebular knots}

\begin{figure}[]
\centering
\includegraphics[width=\hsize]{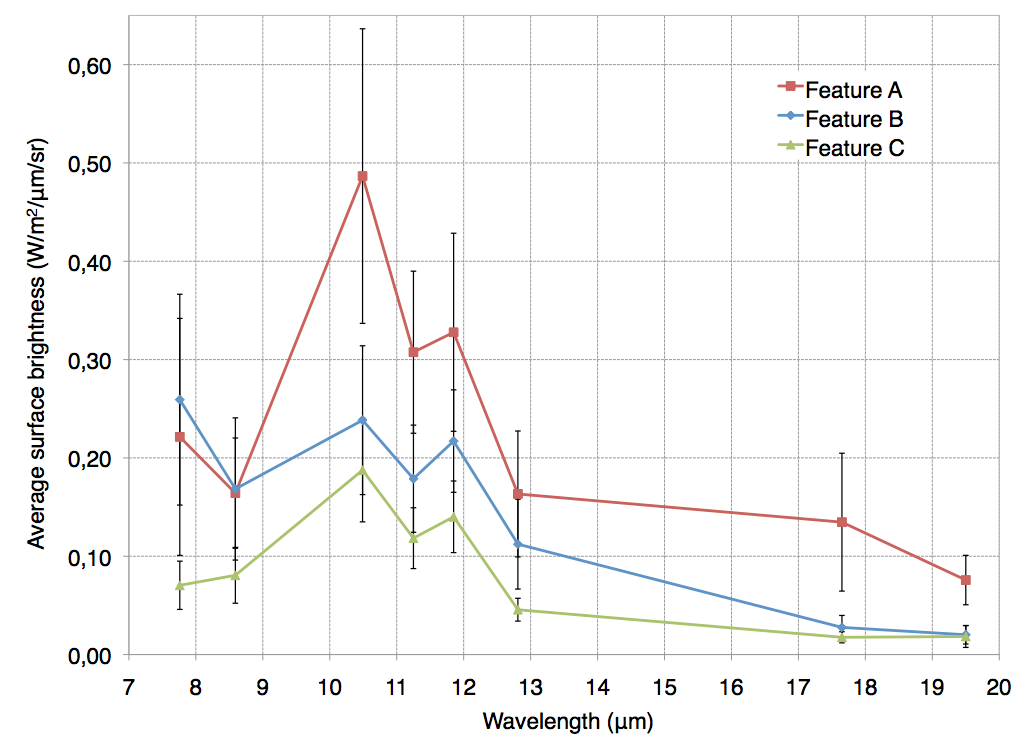}
\includegraphics[width=\hsize]{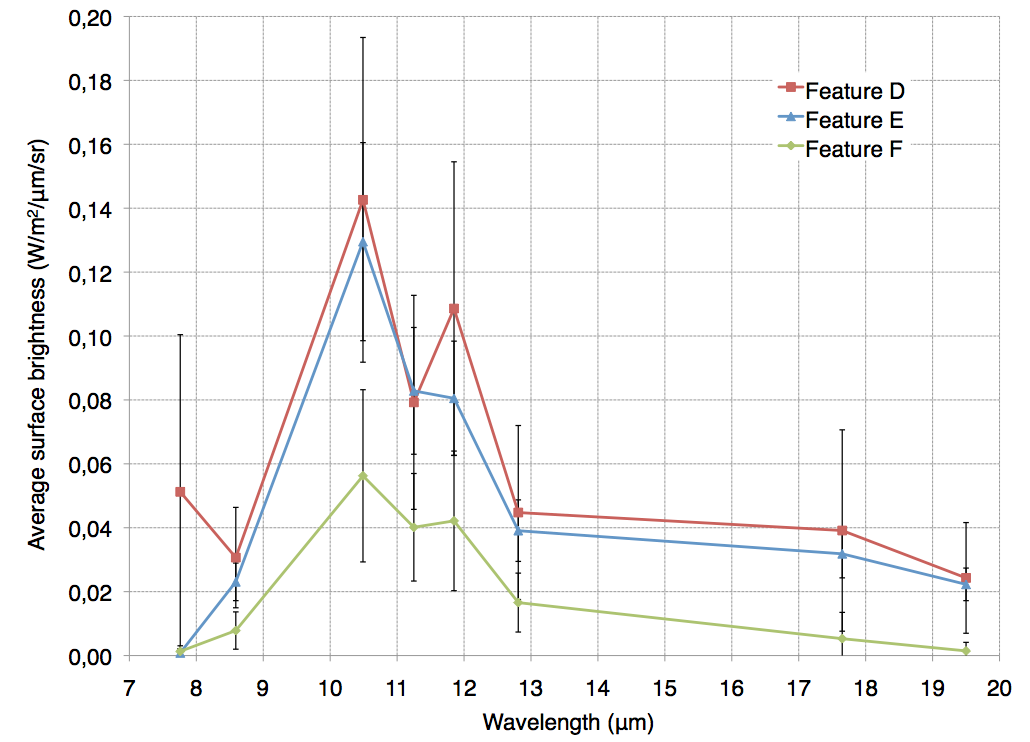}
\caption{Average surface brightness of Features A, B, C (top) and D, E, F (bottom) as defined in Fig.~\ref{Features}, measured over a disk with a diameter of $0.38\arcsec$. The error bars correspond to the standard deviation of the surface brightness over the measurement area, added quadratically to the 20\% systematic calibration uncertainty (for the $N$ band only, see text for details). \label{FeaturesABCDEF}}
\end{figure}

The average surface brightness of the six knots identified in the nebula are presented in Fig.~\ref{FeaturesABCDEF}. We measured the surface brightness over a circular aperture of $0.38\arcsec$ in diameter, a size that corresponds to the angular resolution of VISIR around 13\,$\mu$m. This wavelength is around the center of our full wavelength range, and combined with the image deconvolution we applied, this relatively large aperture results in surface brightness measurements that are largely insensitive to the changing angular resolution with wavelength. As discussed in Sect.~\ref{OverallNebula}, we estimate the systematic uncertainty on the photometry of the nebula to $20\%$. The error bars represented in Fig.~\ref{FeaturesABCDEF} correspond to the dispersion of the flux on each of the measurement apertures, quadratically added to the systematic uncertainty.

\section{Discussion \label{discussion}}

\subsection{Evolutionary time scales\label{evolution}}

An important aspect to consider in any study of Betelgeuse is its variability with time. This makes the comparison of different studies obtained at different epochs more difficult, but it is also a powerful means to sample the physics of the star, especially since the advent of realistic 3D hydrodynamical simulation of supergiant stars (Freytag~\cite{freytag02}, Freytag \& H{\"o}fner~\cite{freytag08}, Chiavassa et al.~\cite{chiavassa10}). Betelgeuse exhibits a variability over two periods of approximately 200 and 2100\,days (Stothers~\cite{stothers10}), by a few tenths of magnitude in visible photometry, and a few km\,s$^{-1}$ in radial velocity. From spectroscopic observations in the visible domain, Gray~(\cite{gray08}) report a characteristic evolution timescale of the velocity variations at the surface of 400\,days, and a range of velocities of $\approx 9$\,km\,s$^{-1}$.

Convective-related surface structures are characterized by two characteristic time scales (Chiavassa et al.~\cite{chiavassa11}): (1) large convective cells, with a size comparable to the stellar radius, that evolve on time scale of years; these convective cells are visible in the infrared. On the top of these cells, there are small convective structures evolving on time scale of months. (2) In the optical region, large convective cells are still visible but they are hidden by numerous short-lived (a few weeks to a few months) small-scale structures (about 5 to 10\% of the stellar radius). An observational evidence of this long timescale is the probable persistence of the southwestern plume reported in Paper~I, that probably corresponds to the bright spot noticed more than 10~years before in HST images of the chromosphere of Betelgeuse (Uitenbroek et~al. \cite{Uitenbroek98}, see also Gilliland \& Dupree~\cite{gilliland96}).

At higher altitudes in the CSE, we can derive a rough estimation of the evolutionary time scale from the ratio of the linear distances involved divided by the displacement velocity of the gas. The velocity of the CO gas was estimated recently by Smith et~al.~(\cite{smith09}) from spatially resolved spectroscopy at $\lambda = 4.6\,\mu$m. They observed typical velocities of 12\,km\,s$^{-1}$, in very good agreement with the slow gas component detected by  Bernat et~al.~(\cite{bernat79}) for a temperature of $\approx 200$\,K, who additionally report a faster expanding batch of gas (+18\,km\,s$^{-1}$) at T$\approx 70$\,K. To convert angular velocities in linear velocities, we use the revised Hipparcos parallax of Betelgeuse from van~Leeuwen~(\cite{vanleeuwen07}), $\pi = 6.55 \pm 0.83$\,mas, a value compatible with the recent results obtained by Harper et~al.~(\cite{harper08}). Considering this distance of $152 \pm 20$\,pc, the photosphere ($\sim$44~mas in the $K$ band, Perrin et~al.~\cite{perrin04}) has a linear radius of 3.4\,AU, and the shell (angular radius of $0.9\arcsec$) is located at $\approx 134$\,AU. Considering a gas velocity of 10-20\,km\,s$^{-1}$, the gas will take $\approx 30$-60\,years to migrate from the photosphere to the inner radius of the dust shell.

In addition to a physical displacement of matter, the change in illumination of the nebula by the evolving photosphere of the star can result in an apparent morphological change of the nebula, as proposed by Skinner et al.~(\cite{skinner97}). This evolution will happen with a typical timescale comparable to the different convective structures present on the star's surface (months to years).
Using the available instrumentation, it therefore appears possible to monitor the evolution of the photosphere of Betelgeuse over a few months using interferometry, and of its CSE over a few years using AO for the inner gaseous envelope, and thermal infrared imaging for the dusty envelope. 

\subsection{Comparison with NACO images}

\begin{figure*}[]
\sidecaption
\includegraphics[width=12.0cm]{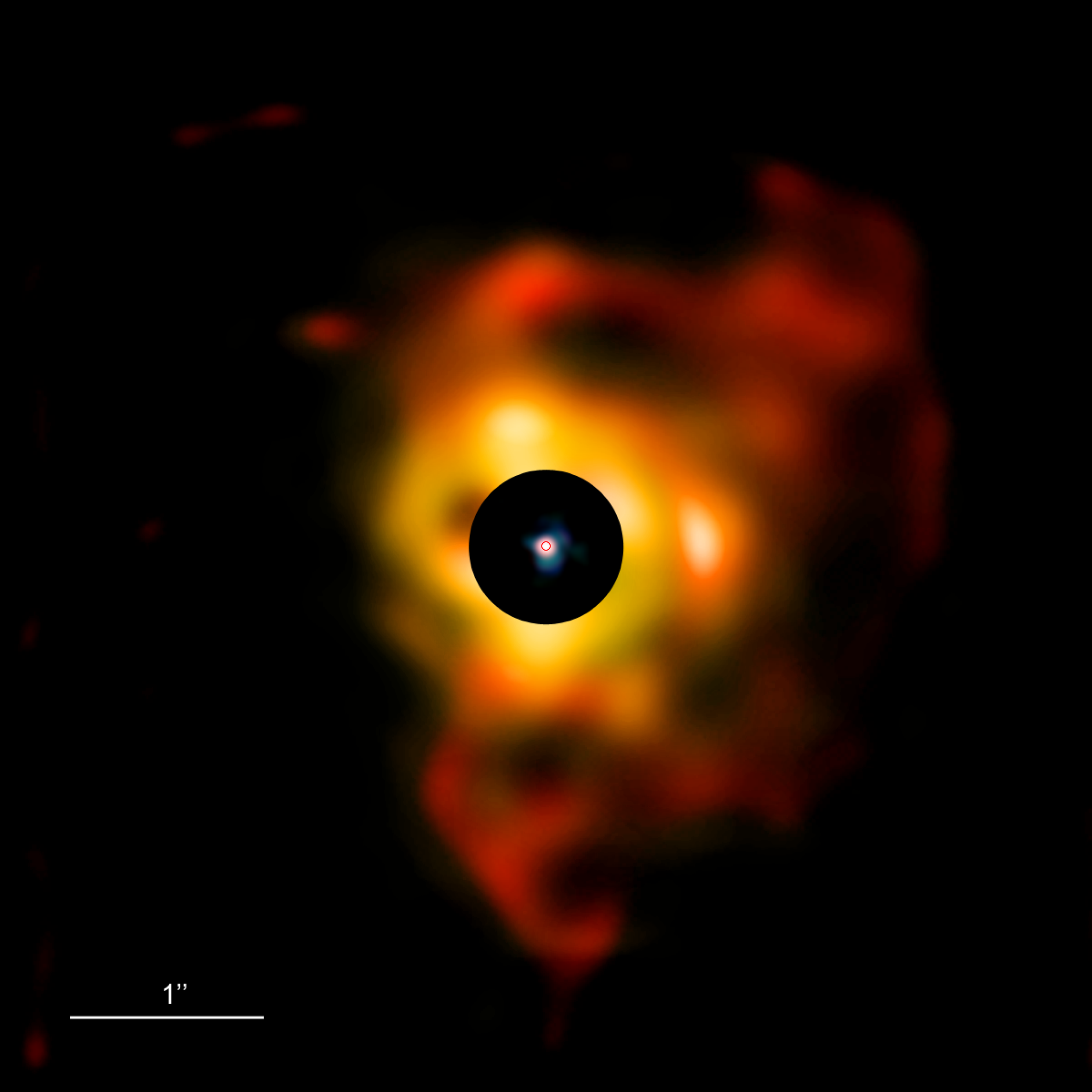}
\caption{Color composite image of the CSE of Betelgeuse. The NACO observations obtained in January 2009 (Paper~I) are reproduced in the central disk, with J coded as blue, H as green and K as red. The apparent angular extension of the near-infrared photosphere ($\theta_{\rm LD} \approx 44$\,mas) is represented with a red circle at the center of the image. The field of view is $5.63\arcsec \times 5.63\arcsec$, with North up and East to the left. The thermal infrared VISIR images reported in the present work are presented with PAH1 coded as yellow, SIV as orange, and Q1 as red. \label{ColorComposite}}
\end{figure*}

A color composite image of the present VISIR observations and the near-infrared observations in the JHK bands (Paper~I, shown in the central disk) is presented in Fig.~\ref{ColorComposite}. The two imaging data sets are represented to scale, the central white spot corresponding to the photosphere of the star. 
One of the key questions we wish to address in our observing program of Betelgeuse is whether CSE asymmetries (plumes in the near-IR, nebular features in the thermal IR) are linked to the convective cells imaged by interferometry at near-infrared wavelengths (Haubois et~al.~\cite{haubois09}, Ohnaka et~al.~\cite{ohnaka09}, Ohnaka et al.~\cite{ohnaka11}, see also Chiavassa et al.~\cite{chiavassa09}).
From the arguments presented in Sect.~\ref{evolution}, it is justified to directly compare the observations obtained with NACO (obtained in January 2009) with the present VISIR images (obtained in November 2010), as the evolutionary time scale of the dusty envelope is significantly longer than the two years separating the two data sets.

Considering their direction and opening angle, the southern and northwestern near-infrared plumes extending from the photosphere correspond well with the thermal infrared filamentary structures labeled \#2 and 3 in Fig.~\ref{Features}.
This is an indication that the molecular flow from the photosphere of the star extends through and feeds the dust shell. The faint loop observed at larger distance from the star (\#6 in Fig.~\ref{Features}) could correspond to a past episode of enhanced mass loss. From the spectral energy distribution of the nebula, it is clear that the shell contains a significant ammount of O-rich dust (silicates, alumina,...), which is expected for Betelgeuse (see e.g. Verhoelst et~al.~\cite{verhoelst09}). This type of dust has strong signatures around 9-10\,$\mu$m, that are visible in the ISO spectrum presented in Fig.~\ref{TotalFlux} as well as in our photometry of the nebula (Fig.~\ref{TotalFlux}). It is expected to form at a few $0.1\arcsec$ above the stellar surface, which also corresponds well to the radius of the shell measured in the VISIR images.

\subsection{Previous observations of Betelgeuse in thermal and far infrared}

The dusty environment of Betelgeuse has been studied at thermal infrared wavelengths using several complementary techniques: interferometry (at a few tens of milliarcsecond scale), large scale imaging (a few arcsecond scale) and spectroscopy.

Interferometric observations at 11\,$\mu$m with the Infrared Spatial Interferometer (ISI) allowed Bester et~al.~(\cite{bester91}) to measure the inner radius of the dust shell at $0.90 \pm 0.05\arcsec$, which was confirmed by Danchi et~al.~(\cite{danchi94}). This radius is in very good agreement with the shell we observe with VISIR (Fig.~\ref{Features}). Although circular symmetry is always hypothesized, it appears in our images that this shell is not complete nor uniform. Townes et~al.~(\cite{townes09}) reported a decrease of the angular diameter of Betelgeuse by 15\% between 1993 and 2009, from interferometric measurements (also with ISI), using two-telescope (up to 2002) and 3-telescope (2002-2009) configurations. While the changing fringe visibility may indeed be related to a change in the photospheric size of the star, another explanation could be that the complex structure of the CSE affected the measurement. Its high brightness could induce a significant variation of the interferometric visibility with baseline length and orientation, that would result in a deviation from the single star model, and consequently a bias on the derived angular diameters. Tatebe et al.~(\cite{tatebe07}) identified an elongation of the disk of Betelgeuse at 11.15\,$\mu$m using ISI with 3 telescopes (their Fig. 2), approximately in the North-South direction. To the North of Betelgeuse, our VISIR images show a bright, jet-like structure (\#1 in Fig.~\ref{Features}). It is clearly visible in the 11.25\,$\mu$m VISIR image (Fig.~\ref{PSF_decLR_subNQ}), and its location well within the field of view of ISI may explain the measured apparent elongation.

An original intermediate approach between long baseline interferometry and direct imaging was chosen by Hinz et~al.~(\cite{hinz98}). Using two mirrors of the Multiple Mirror Telescope separated by a baseline of 5\,m and nulling interferometry, these authors observed the CSE of Betelgeuse at 10\,$\mu$m after cancellation of the flux from the star itself. They determined a triangular shape for the nebula that we also observe in our images, with the same orientation and extension. 
Harper et~al.~(\cite{harper09}) observed Betelgeuse using the TEXES spectrograph installed on NASA's IRTF, and identified narrow iron emission lines at 17.94 and 24.52\,$\mu$m, that form close to the stellar surface (within a few stellar radii). Such emission could explain at least part of the unresolved emission we observe in the Q1 and Q3 VISIR images at the star's location (Fig.~\ref{PSF_decLR_subNQ})

Few direct imaging observations of Betelgeuse in the thermal infrared domain are present in the literature. Skinner et al.~(\cite{skinner97}) obtained images of Betelgeuse at 8.2, 9.7 and 12.5\,$\mu$m using UKIRT. They resolved the circumstellar nebula, and noticed a significant evolution over a period of one year. They attribute this evolution to a change in illumination of the nebula, caused by the presence of variable spots on the star's surface. From a model that includes also radio observations with the VLA (see also the preceding modeling work by Skinner \& Whitmore~\cite{skinner87}), they determine an internal radius of the dust shell of $0.5\arcsec$, which is slightly smaller than the shell we observe with VISIR. Direct imaging of Betelgeuse in the thermal infrared was also obtained by Rinehart et~al.~(\cite{rinehart98}) at 11.7 and 17.9\,$\mu$m using the Hale 5\,m telescope. From a model fit to their images, they derive an inner dust shell radius of $1.0 \pm 0.1\arcsec$, also in excellent agreement with our observations. A very similar radius for dust formation ($\approx$33\,R$_* = 0.73\arcsec$) was also proposed by Harper et al.~(\cite{harper01}) from a model of the extended atmosphere of the star.

At a much larger spatial scale, the interface between the wind of Betelgeuse and the interstellar medium was recently observed in the far infrared (65 to 160\,$\mu$m) by Ueta et al.~(\cite{ueta08}) using the AKARI satellite. This bow shock is located at a radius of $4.8\arcmin$ from the star, corresponding to $\approx100$\,times the extension of the dusty envelope we observe with VISIR.
 
\subsection{A stellar companion ?}

Some interesting structures were detected by Smith et~al.~(\cite{smith09}) $\approx 1\arcsec$ west of the star, which corresponds to the position of the bright knot A (Fig.~\ref{Features}). Its spectral energy distribution (Fig.~\ref{FeaturesABCDEF}) indicates that it contains O-rich dust. Its compact aspect and relatively high brightness are however intriguing, and it is not excluded that it may correspond to a stellar companion of Betelgeuse. Such a companion would be in a position to accrete material lost by the supergiant, and therefore present dust and molecular emission features. Considering the mass of Betelgeuse ($\approx 10-20\,M_\odot$), its {\it a priori} probability to be part of a binary or multiple system is relatively high ($\approx 70$\%, see e.g. Kouwenhoven et al.~\cite{kouwenhoven07}). It should be noted that knot A is located outside the field of view of the NACO observations reported in Paper~I, and we cannot confirm this hypothesis with these observations.

The presence of two companions in orbit around Betelgeuse was proposed by Karovska et~al.~(\cite{karovska86}) based on speckle interferometric observations obtained at epoch 1983.88, but they were not confirmed. One of the sources detected by these authors was found at a position angle of $278 \pm 5^\circ$ and a radius of $0.51 \pm 0.01\arcsec$ from Betelgeuse, i.e. closer but at a similar position angle compared to our knot A (Table~\ref{feature_coords} and Fig.~\ref{Features}). If we consider that it is the same source observed at two epochs, its angular velocity projected on the plane of the sky would be 33\,mas\,yr$^{-1}$ (VISIR epoch is 2010.86). At the distance of Betelgeuse, this angular velocity corresponds to a projected linear velocity of $\approx 24$\,km\,s$^{-1}$ relatively to Betelgeuse. This figure is comparable to the typical velocities derived from CO lines, considering the different projection effect (spectroscopy measures the velocity perpendicular to the plane of the sky, while imaging gives measures it in the plane of the sky).  However, this apparent displacement appears too slow for the orbital motion of a close-in companion of Betelgeuse, that would have a typical period of a few years at a separation of $\approx 1\arcsec$. The possibility nevertheless remains that knot~A corresponds to a relatively distant companion (thus with a slower orbital motion), whose apparent proximity to the star would be caused by projection effects.

While this hypothesis cannot be formally excluded from our data, the possibility that knot A is associated with the presence of a stellar companion of Betelgeuse appears unlikely. The most probable hypothesis is that it is simply an overdensity of the dust in the CSE, that may have been observed in visible scattered light by Karovska et al.~(\cite{karovska86}).

\section{Conclusion \label{conclusion}}

Our VISIR observations of the CSE of Betelgeuse show the presence of a complex and bright nebula. The observed brightening of the nebula for wavelengths greater than $\approx$9-10\,$\mu$m points at the presence of O-rich dust, such as silicates or alumina. The shell that we observe at a radius of $0.9\arcsec$ presents a similar spectral energy distribution as the rest of the nebula, and could correspond to the condensation radius of certain dust species, or to a past episode of enhanced mass loss. A promising correspondance is observed between the plumes reported in Paper~I from NACO observations and the VISIR nebular features. The knots and filamentary extensions of the nebula observed at larger distances from Betelgeuse appear to correspond to inhomogeneities in the mass lost by the star in the recent past, probably within the last few centuries. Further observations are expected to clarify the nature and composition of the nebular features identified in our images, using spatially resolved spectroscopy of the CSE.

\begin{acknowledgements}
We are grateful to ESO's Director General Prof. Tim de Zeeuw for the allocation of observing time to our program, as well as to the VISIR instrument team, in particular Dr. Yazan Al Momany and Dr. Mario van den Ancker, for the successful execution of the observations. STR acknowledges partial support from NASA grant NNH09AK731.
We thank the referee, Dr. Peter Tuthill, for his suggestions that led to improvements of this article.
This research received the support of PHASE, the high angular resolution partnership between ONERA, Observatoire de Paris, CNRS and University Denis Diderot Paris 7.
It is based in part on observations with ISO, an ESA project with instruments funded by ESA Member States (PI countries: France, Germany, the Netherlands and the United Kingdom) and with the participation of ISAS and NASA.
We used the SIMBAD and VIZIER databases at the CDS, Strasbourg (France), and NASA's Astrophysics Data System Bibliographic Services.
\end{acknowledgements}


\end{document}